\documentclass[12pt,aps,prd,preprint,tightenlines,
amsmath,amssymb,showpacs,nofootinbib
]{revtex4}

\usepackage{amsmath}         
\usepackage{amssymb}         
\usepackage{amsfonts}        
\usepackage{graphicx}        

\usepackage{lipsum}        
\usepackage{color}         
\usepackage{enumerate}		%

\graphicspath{{figures/}}	

\renewcommand{\vec}[1]{\mathbf{#1}} 
\renewcommand{\email}[1]{\href{mailto:#1}{\texttt{#1}}}

\let\oldenumerate\enumerate
\renewcommand{\enumerate}{
  \oldenumerate
  \setlength{\itemsep}{1pt}
  \setlength{\parskip}{0pt}
  \setlength{\parsep}{0pt}
}

\let\olditemize\itemize
\renewcommand{\itemize}{
  \olditemize
  \setlength{\itemsep}{1pt}
  \setlength{\parskip}{0pt}
  \setlength{\parsep}{0pt}
}

\newcommand{\PRE}[1]{{#1}}  

\newcommand{\mev}{\text{MeV}}
\newcommand{\gev}{\text{GeV}}
\newcommand{\tev}{\text{TeV}}

\newcommand{\cm}{\text{cm}}
\newcommand{\m}{\text{m}}
\newcommand{\km}{\text{km}}

\newcommand{\sr}{\text{sr}}
\newcommand{\s}{\text{s}}
\newcommand{\yr}{\text{yr}}

\renewcommand{\eqref}[1]{Eq.~(\ref{#1})}

\newcommand{\secref}[1]{Sec.~\ref{sec:#1}}

\newcommand{\figref}[1]{Fig.~\ref{fig:#1}}

\newcommand{\Pdet}{P_{\text{det}}}

\usepackage[colorlinks=true,citecolor=blue,linkcolor=blue,
urlcolor=blue,hypertexnames=false]{hyperref}

\usepackage{microtype}				

\begin{document}

\preprint{UCI-TR-2016-02}

\title{\PRE{\vspace*{1.5in}}
{\Large Dark Sunshine: Detecting Dark Matter \\ 
through Dark Photons from the Sun}
\PRE{\vspace*{.5in}}}

\author{Jonathan L.~Feng\footnote{\email{jlf@uci.edu}}
}
\affiliation{Department of Physics and Astronomy, University of
  California, Irvine, California 92697, USA
\PRE{\vspace*{.4in}}
}

\author{Jordan Smolinsky\footnote{\email{jsmolins@uci.edu}}
}
\affiliation{Department of Physics and Astronomy, University of
  California, Irvine, California 92697, USA 
\PRE{\vspace*{.4in}}
}

\author{Philip Tanedo\footnote{\email{flip.tanedo@uci.edu}}
}
\affiliation{Department of Physics and Astronomy, University of
  California, Irvine, California 92697, USA
\PRE{\vspace*{.4in}}
}


\begin{abstract}
\PRE{\vspace*{.2in}}
\noindent
Dark matter may interact with the Standard Model through the kinetic mixing of dark photons, $A'$, with Standard Model photons. 
Such dark matter will accumulate in the Sun and annihilate into dark photons.  The dark photons may then leave the Sun and decay into pairs of charged Standard Model particles that can be detected by the Alpha Magnetic Spectrometer. 
The directionality of this ``dark sunshine'' is distinct from all astrophysical backgrounds, providing an opportunity for unambiguous dark matter discovery by AMS. 
We perform a complete analysis of this scenario including Sommerfeld enhancements of dark matter annihilation and the effect of the Sun's magnetic field on the signal, and we define a set of cuts to optimize the signal probability.
With the three years of data already collected, AMS may discover dark matter with mass $1~\tev \alt m_X \alt 10~\tev$, dark photon masses $m_{A'} \sim \mathcal O(100)$~MeV, and kinetic mixing parameters $10^{-10} \alt \varepsilon \alt 10^{-8}$. 
The proposed search extends beyond existing beam dump and supernova bounds, and it is complementary to direct detection, probing the same region of parameter space.
 
 \end{abstract} 


\maketitle

\section{Introduction}

One of the clear signs for physics beyond the Standard Model (SM) is the existence of dark matter. The correct present-day abundance of dark matter can be realized if dark matter with weak-scale mass annihilates into SM particles with approximately weak-interaction couplings.
This framework implies promising direct, indirect, and collider searches for dark matter, but this promise is not generic to all dark matter candidates.  One limit in which thermal relic dark matter may hide from experimental searches is if it interacts through a light mediator. In this case, annihilation into on-shell mediators may occur with the correct couplings for a thermal relic, but direct detection and collider bounds can be parametrically suppressed by the mediator coupling to the SM.  

A simple realization of this scenario is a dark sector with a broken U(1) gauge symmetry, which provides a massive ``dark photon''~\cite{Kobzarev:1966qya, Okun:1982xi}. This dark photon may kinetically mix with the SM photon with a very small mixing parameter if, for example, the mixing is produced by loops of heavy particles~\cite{Holdom:1985ag,Holdom:1986eq}. 

This type of dark sector predicts a novel class of indirect detection signals. In this framework, dark matter is captured by large gravitating bodies and annihilates into dark photons. The decay products of these dark photons can be detected if they escape the  gravitating object. The formalism for dark matter capture and annihilation was developed many years ago for the case of dark matter annihilating in the Sun or Earth to neutrinos~\cite{Freese:1985qw,Press:1985ug,Silk:1985ax,  Krauss:1985aaa,Griest:1986yu,Gaisser:1986ha,Gould:1987ju, Gould:1987ir,1988ApJ...328..919G,Gould:1991hx}. In the past few years, studies have begun to explore the case of annihilation into new, light SM singlet particles~\cite{Batell:2009zp}, including the specific case of dark photons~\cite{Schuster:2009au,Schuster:2009fc,Meade:2009mu}.

In Ref.~\cite{Feng:2015hja}, we carried out a detailed examination of dark matter annihilating to dark photons in the center of the Earth.  We found that this could result in spectacular signals in the IceCube Neutrino Observatory and possibly also space-based detectors such as the Fermi Large Area Telescope (LAT) and the Alpha Magnetic Spectrometer (AMS). As an example, in currently unconstrained regions of parameter space with dark matter masses $100~\gev \alt m_X \alt 10~\tev$, dark photon masses $m_{A'} \sim \mev - \gev$, and kinetic mixing parameters $10^{-10} \alt \varepsilon \alt 10^{-8}$, this scenario predicts up to thousands of TeV-energy $e^+/e^-$, $\mu^+/\mu^-$, and hadron pairs from the center of the Earth streaming through the IceCube detector each year.  Experimental searches for this signal will therefore either exclude new regions of parameter space or provide the first unambiguous signal of dark matter.  Additionally, in contrast to the standard case of indirect detection of neutrinos, in the dark photon case, all of the annihilation products from a single dark matter particle can be detected, allowing one to reconstruct the dark matter mass from a few clean signal events. 

\begin{figure}[tb]
\includegraphics[width=.99\linewidth]{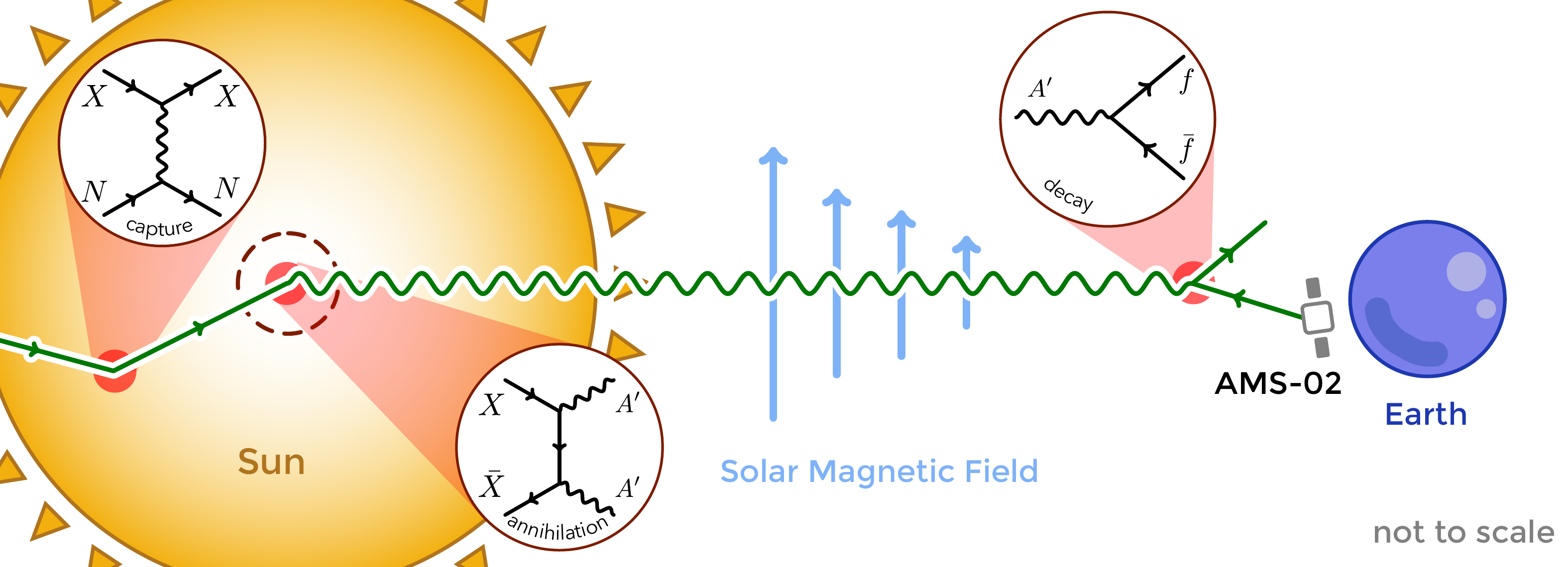}
\vspace*{-0.1in}
  \caption{Dark matter is captured by elastic $X N \to X N$ scattering off nuclei, collects in the center of the Sun, and annihilates to dark photons, $XX \to A' A'$.  These dark photons  then leave the Sun and decay to SM particles, including positrons that may be detected by the Alpha Magnetic Spectrometer on the International Space Station.}
  \label{fig:picture}
\vspace*{-0.1in}
\end{figure}

In this work we examine the complementary possibility that dark matter accumulates not in the Earth, but in the Sun, annihilating to dark photons (``dark sunshine'') which decays to SM particles.  The complete process is shown schematically in Fig.~\ref{fig:picture}. 

At first sight, replacing the Earth with the Sun may appear to be a simple substitution, but this is far from the case.  There are some obvious differences: the longer propagation distance required for dark photons to escape the Sun compared to the Earth implies that searches for solar dark photons probe  smaller kinetic mixing parameters $\varepsilon$.  In addition, the Sun's size provides more targets and a bigger gravitational potential to assist dark matter capture, and the solar capture rate is more reliably calculated than the Earth's. 
There is also a very significant new complication, however: for dark photons from the Sun, the magnetic fields of the Sun and Earth deflect the dark photon decay products, potentially ruining the directionality of the signal.  Cuts to define the signal region and optimize the signal above background must, therefore, be very carefully defined.  

In this work, we perform a complete, systematic analysis of solar capture of dark matter and its subsequent annihilation into dark photons. 
We go beyond prior analyses by including the effect of non-perturbative Sommerfeld enhancements of the annihilation rate.  We also consider self-capture in the Sun~\cite{Zentner:2009is}, which we find to be a subleading effect in the regions of parameter space with significant event rates. In addition, we model the effect of the solar magnetic field on the signal. We focus on the reach of AMS-02 to detect the signal in positrons and optimize the signal over background by defining stringent cuts to reduce the background to almost negligible levels.  The analysis makes essential use of AMS's excellent angular resolution~\cite{Gallucci:2015fma}, which has not previously been utilized in its dark matter searches.

Including these effects, we show that AMS can discover dark matter through the dark sunshine signal for parameters 100~GeV $\alt m_X \alt$ 10~TeV, $m_{A'} \sim \mathcal O(100)$~MeV, and $10^{-10} \alt \varepsilon \alt 10^{-8}$. The signal probes a region in parameter space that is unconstrained by beam dump and supernova bounds.  This region is also probed by direct detection, and so this suggested search provides a complementary probe.  
Such values
of $\varepsilon$ are naturally induced, for example, by degenerate
bi-fundamentals in grand unified theories~\cite{Collie:1998ty}.  These
values of $m_{A'}$ and $\varepsilon$ also produce dark matter
self-interactions that have been suggested to solve small scale
structure anomalies~\cite{Tulin:2013teo} and may simultaneously
explain the excess of gamma rays from the galactic center recently
observed by the Fermi Large Area Telescope~\cite{Kaplinghat:2015gha}.

The Fermi-LAT collaboration has set investigated the possibility of dark matter captured in the Sun annihilating into on-shell mediators~\cite{Ajello:2011dq}. The approach taken there was to use light mediators as a general motivation for searches for asymmetries, for example, for an excess of positrons from the hemisphere including the Sun over the opposite hemisphere.  This study is complementary to that work in that we maximize the search reach by defining more stringent cuts that reduce the background to near-negligible levels.  In addition, we consider the dark photon mediator specifically, determine the reach in this model's parameter space, and compare it to the reach of other experimental and observational constraints.

\section{Dark Matter Interactions Through a Dark Photon}

The dark photon $A'$ is the gauge boson of a broken U(1) symmetry that kinetically mixes with the hypercharge boson. The diagonalization of the Hamiltonian from the kinetically mixed gauge--basis states to physical states is detailed in the Appendix. 
When the dark photon mass is very light, the mixing with the $Z$ boson is negligible and this system may be treated as a mixing between the photon and the dark photon. 
The effective Lagrangian for the photon--dark photon system is
\begin{align}
	\mathcal L &=
	-\frac 14 F_{\mu\nu}F^{\mu\nu}
	-\frac 14 F'_{\mu\nu}F'^{\mu\nu}
	+ \frac{1}{2} m^2_{A'} A'^2
	- \sum_f q_f e (A_\mu + \varepsilon A'_\mu) \bar f\gamma^\mu f
	- g_X A'_\mu \bar X\gamma^\mu X \ ,
	\label{eq:Lagrangian}
\end{align}
where we sum over SM fermions $f$ with electric charge $q_f$, $\varepsilon$ is the kinetic mixing parameter in the physical basis, and $g_X$ is the dark U(1) gauge coupling. We present our results in terms of the electromagnetic and dark fine structure constants, $\alpha = e^2/(4\pi)$ and $\alpha_X = g_X^2/(4\pi)$. 

Dark photons decay to SM fermions, $f$, with a branching ratio
\begin{align}
	\Gamma(A'\to f\bar f)
	&=
	\frac{\varepsilon^2 q_f^2 \alpha (m_{A'}^2 + 2m_f^2)}{3m_{A'}}
	\sqrt{1-\frac{4m_f^2}{m_{A'}^2}} \ .
\end{align}
In the limit where $m_X \gg m_{A'} \gg m_e$,  the dark photon decay length is
\begin{align}
	L
	&= 
	R_\odot \, \text{Br}(A'\to e^+e^-)
	\left( \frac{1.1 \times 10^{-9}}{\varepsilon} \right)^2
	\left( \frac{m_X/m_{A'}}{1000} \right)
	\left( \frac{100\text{ MeV}}{m_{A'}} \right),
	\label{eq:decay:length}
\end{align}
where $R_\odot = 7.0 \times 10^{10}\text{ cm} = 4.6 \times 10^{-3}\text{ au}$ is the radius of the Sun, and the branching ratio to $e^+e^-$ can be determined from hadron production at colliders and is between 40\% and 100\% for the range 1~MeV $\alt m_{A'} \alt$ 500~MeV~\cite{Buschmann:2015awa}. 

We consider two choices for the dark photon couplings. First, we consider the case where the present dark matter abundance is set by thermal freeze out with respect to the annihilation process $X\bar X \to A' A'$. The Born approximation cross section valid at freeze out is~\cite{Liu:2014cma}
\begin{align}
\langle \sigma_{\text{ann}} v \rangle^\text{Born} = 
\frac{\pi \alpha_X^2}{m_X^2} 
\frac{\left(1 - m_{A'}^2/m_X^2\right)^{3/2}} {\left[1-m_{A'}^2 / (2 m_X^2) \right]^2} 
\ .
\label{eq:annihilation:rate:Born}
\end{align}
Obtaining the observed $\Omega_X h^2 = 0.12$ from thermal freeze out requires $\langle \sigma_\text{ann} v \rangle = 2.2 \times 10^{-26} \text{ cm}^3/\text{s}$~\cite{Steigman:2012nb}, so that
\begin{align}
	\alpha_X^{\text{th}} &= 0.035 \left(\frac{m_X}{\text{TeV}}\right) \ .
	\label{eq:alphaX:th}
\end{align}
Alternatively, one may assume that the dark matter abundance is set by non-thermal dynamics and allow $\alpha_X$ to take its maximal experimentally-allowed value.  The most stringent bounds come from the imprint of dark matter annihilation products on the cosmic microwave background (CMB)~\cite{Adams:1998nr,Chen:2003gz,Padmanabhan:2005es,Slatyer:2015jla}. We fit the results of Ref.~\cite{Slatyer:2015jla} and find that the maximum coupling allowed by the CMB is
\begin{align}
	\alpha_X^\text{max} = 0.17 \left(\frac{m_X}{\text{TeV}}\right)^{1.61} 
	\label{eq:alphaX:max}
\end{align}
in the range of phenomenologically relevant masses.

\section{Experimental Bounds on Dark Photon Mediators}
\label{sec:bounds}

Here we briefly review the bounds on dark photons that are most relevant in the parameter space relevant to this work. 

\subsection{Direct Detection}

Direct detection experiments bound dark photon mediated interactions with weak-scale dark matter. These were recently examined in Ref.~\cite{DelNobile:2015uua}, which highlighted that the exclusion contour in the ($m_A$, $\varepsilon$) plane becomes independent of $m_X$ for small $m_A$ when the contact-interaction limit breaks down. This is easy to understand: in the $m_A \ll m_X$ limit, the $X$--nucleon cross section and annihilation rate scale as
\begin{align}
\sigma_{Xn} &\sim \alpha_X \rho_0 \sim \frac{\alpha_X}{m_X}
&
\langle\sigma_\text{ann}v\rangle &\sim  \frac{\alpha_X^2}{m_X^2} \ .
\end{align}
Fixing $\alpha_X$ to yield the thermal relic cross section $\langle \sigma_\text{ann} v\rangle = 2.2\times 10^{-26}\text{ cm}^3/\text{s}$ gives $\alpha_X \sim m_X$, and so the direct detection bounds are constant in $m_X$ for a thermal relic.

\subsection{Colliders and Fixed-Target Experiments}

Direct searches for dark photons production at colliders and beam dump experiments are reviewed in Ref.~\cite{Essig:2013lka}. These searches do not make use of the dark photon--dark matter coupling and can thus be plotted in the $(m_{A'},\varepsilon)$ plane independently of the dark matter mass. In the mass range probed by this study, $\text{MeV} < \m_{A'} < \text{GeV}$ and $10^{-12} < \varepsilon < 10^{-7}$, the most relevant bounds are from the E137 beam dump experiment~\cite{Bjorken:1988as, Bjorken:2009mm}, 
the LSND neutrino experiment~\cite{Batell:2009di, Essig:2010gu, PhysRevC.58.2489}, and the CHARM fixed target experiment~\cite{Gninenko2012244, Bergsma:1986is}. For the dark matter mass range where AMS is sensitive to solar dark photons, these collider experiments are less sensitive than the direct detection bounds from LUX presented in Ref.~\cite{DelNobile:2015uua}.

\subsection{Indirect Detection}

Bounds on dark matter annihilation into dark photons in the present day coming from the diffuse positron spectrum constrain the dark sector coupling, $\alpha_X$~\cite{Meade:2009mu,Schuster:2009au,Batell:2009zp}. These bounds do not reach the thermal coupling and are weaker than the CMB bounds that define our maximal coupling in Eq.~(\ref{eq:alphaX:th}).

\subsection{Supernova Bounds}

Independent of the dark matter properties, light mediators are constrained by the cooling of supernova by mediator emission~\cite{Dent:2012mx, Dreiner:2013mua, Essig:2013lka, Kazanas:2014mca, Rrapaj:2015wgs,Mahoney:2017jqk}. In particular, Ref.~\cite{Mahoney:2017jqk} recently refined the analysis of supernova cooling and found that the bounds on dark photons are nearly an order of magnitude weaker than previously published limits. Separately, the absence of a prompt MeV $\gamma$-ray signal from supernova 1987A sets additional bounds on the $(\varepsilon, m_{A'})$ plane~\cite{Kazanas:2014mca}. Ref.~\cite{Zhang:2014wra} pointed out that dark matter interactions may weaken these bounds when the dark matter is light $(m_X \alt \text{GeV})$. 

\subsection{Cosmology}

The cosmic microwave background sets bounds on dark matter annihilation products in the early universe~\cite{Adams:1998nr,Chen:2003gz,Padmanabhan:2005es}.
In addition to the CMB bounds from Ref.~\cite{Slatyer:2015jla} that set the maximum phenomenologically allowed $\alpha_X$ in \eqref{eq:alphaX:max}, the impact of late dark photon decays on big bang nucleosynthesis and the CMB constrains the $(m_{A'}, \varepsilon)$ plane for $m_{A'} \alt \text{GeV}$~\cite{Fradette:2014sza}.

\section{Dark Matter Accumulation in the Sun}

Dark matter is captured in the Sun if elastic collisions with solar nuclei transfer enough energy that the dark matter's velocity falls below the Sun's escape velocity. Dark matter may also be self-captured by scattering off of already-captured dark matter~\cite{Zentner:2009is}. The captured dark matter accumulates in the solar core and thermalizes. This accumulation is balanced by annihilation into pairs of dark photons. Due to the low temperature at the core of the Sun, this annihilation rate is Sommerfeld enhanced from dark photon-mediated interactions at low relative velocity. 

The number of dark matter particles in the Sun, $N_X$, satisfies the rate equation
\begin{align}
	\dot N_X = C_\text{cap} + C_\text{self} N_X - C_\text{ann} N_X^2 \ ,
	\label{eq:rate}
\end{align}
where the $C$ coefficients encode the capture rate, self-capture rate, and annihilation rate. We ignore the effect of dark matter evaporation, which is negligible for dark matter masses above $\mathcal O(10)$~GeV~\cite{Griest:1986yu,Gaisser:1986ha}. The equilibrium time scale for this expression is
\begin{align}
	\tau = \frac{1}{\sqrt{C_\text{cap}C_\text{ann} + \frac 14 C_\text{self}^2}} \ .
	\label{eq:tau}
\end{align}

Below we show that the self-capture effect on the equilibrium time is negligible for our parameter range of interest. The solution to the rate equation in the relevant limit $C_\text{self}^2 \ll C_\text{cap}C_\text{ann}$ is
\begin{align}
	N_X &= \sqrt{\frac{C_\text{cap}}{C_\text{ann}}} \tanh \frac{t}{\tau}
	&
	\Gamma_\text{ann} &\equiv
	\frac 12 C_\text{ann} N_X^2
	= \frac 12 C_\text{cap} \tanh^2 \frac{t}{\tau} \ .
	\label{eq:NX:Cann}
\end{align}
The factor of $1/2$ accounts for the fact that two dark matter particles are removed in each annihilation.
When the age of the Sun is greater than the equilibrium time, $\tau_\odot \simeq 4.5~\text{Gyr} > \tau$, the Sun is saturated with dark matter and the annihilation rate is maximized and matches the accumulation rate. For $\tau_\odot < \tau$, the dark matter population in the Sun is still growing and the $\tanh^2(\tau_\odot/\tau)$ factor suppresses the annihilation rate relative to the capture rate. 

We now examine each term in \eqref{eq:rate}.

\subsection{Dark Matter Capture}

The capture rate for dark matter scattering off of a particular nuclear species $N$ in the Sun is the integral of the differential cross section over the volume of the Sun; the incident dark matter velocity, $w$; and the nuclear recoil energies, $E_R$, for which capture occurs:
\begin{align}
C_{\text{cap}}^N 
&= 
n_X 
\int_0^{R_{\odot}} dr \, 4\pi r^2 n_N(r) 
\int_0^{\infty} dw\, 4 \pi w^3 f_{\odot}(w, r) 
\int
\left. dE_R \frac{d\sigma_N}{d E_R}  \right|_{\text{capture}}
\ ,
\label{eq:Ccap:N:general}
\end{align}
where $n_X = \rho_X/m_X$ is the local dark matter number density, $n_N(r)$ is the $N$ number density at a distance $r$ from the solar center, $f_\odot(w,r)$ is the dark matter velocity distribution at that position, and $d\sigma_N/dE_R$ is the elastic scattering cross section. The full capture rate is the sum over all nuclear species in the Sun, $C_\text{cap} = \sum_N C_\text{cap}^N$. 

The velocity of dark matter asymptotically far from the Sun, $u$, is distributed according to a Maxwell--Boltzmann-like velocity distribution. In the neighborhood of the Sun, this distribution is distorted due to the solar gravitational potential. Taking this acceleration into account and invoking energy conservation, the incoming dark matter velocity $w$ for an interaction with a nucleus in the Sun is
\begin{align}
	w^2 = u^2 + v_\odot^2(r) \ ,
\end{align}
where $v_\odot(r)$ is the escape velocity at a distance $r$ from the solar center.
The dark matter velocity distribution,
$f_\odot(w,r)$, thus satisfies
\begin{align}
	w^3 f_\odot(w,r)\, dw &= u\left[u^2 + v_\odot^2(r)\right] f(u) \, du \ .
	\label{eq:w3:f:dw}
\end{align}
This may then be substituted directly into Eq.~(\ref{eq:Ccap:N:general}). We use the asymptotic velocity distribution in the solar rest frame,
\begin{align}
	f_\odot(u) &= 
	\frac 12 
	\int_{-1}^1 dc
	\, f\left(
	\sqrt{
	u^2 + u_\odot^2 + 2 u u_\odot c
	}
	\right) ,
\end{align}
where $u_\odot= 233\text{ km/s}$ is the solar velocity in the galactic rest frame, and 
\begin{align}
	f(u) &= N\left[
	\exp\left( \frac{v_\text{gal}^2 - u^2}{ku_0^2} \right)-1
	\right]^k\Theta(v_\text{gal}-u) \ ,
\end{align}
where $v_\text{gal}$ is the galactic escape velocity and $N$ is chosen to normalize the distribution to integrate to unity. The Maxwell--Boltzmann distribution is recovered for $k=0$ and  $v_\text{gal}\to \infty$. The astrophysically favored range of parameters is~\cite{Baratella:2013fya} 
\begin{align}
220 \text{ km/s} < u_0 &< 270 \text{ km/s}
&
450 \text{ km/s} < v_\text{gal} &< 650\text{ km/s}
&
1.5 < k < 3.5 \ .
\end{align}
In this analysis we use the central values of these ranges. We confirm that varying these parameters in this range does not perceptibly alter the dark matter capture rate in the Sun~\cite{Choi:2013eda}.

The differential elastic scattering cross section in the non-relativistic limit is
\begin{align}
	\frac{d\sigma_N}{dE_R}
	&=
	8\pi \varepsilon^2 \alpha_X \alpha Z_N^2
	\frac{m_N}{w^2(2m_N E_R + m_{A'}^2)^2} |F_N|^2 \ ,
	\label{eq:dsig:dEr}
\end{align}
where the Helm form factor is
\begin{align}
|F_N|^2 &= \exp \left(- \frac{E_R}{E_N} \right)
	&
	E_N &= \frac{0.114~\gev}{A_N^{5/3}}
	\ ,
\end{align}
for a target nucleus $N$ with mass $m_N$ and atomic number $A_N$. 
Dark matter captures in the Sun when the outgoing $X$ velocity is less than the escape velocity $v_\odot(r)$ at distance $r$ from the solar center. 
This occurs if sufficient energy, $E_R$, is transferred to the nucleus.
The minimum energy transfer from an incident $X$ with velocity $w$ to the nucleus at distance $r$ from the Sun in order for the dark matter to be captured is
\begin{align}
	E_\text{min} = \frac 12 m_X \left[w^2 - v_\odot^2(r)\right].
\end{align}
The range of allowed recoil energies is determined by kinematics. Writing the dark matter--nucleus reduced mass as $\mu_N$, the lab frame recoil energy is
\begin{align}
	E_R &= 
	\frac 12 E_\text{max}(1-\cos\theta_\text{CM})
	&
	E_\text{max} &= \frac{2\mu_N^2 w^2}{m_N} \ ,
\end{align}
where we have identified the maximum kinematically permitted recoil energy, $E_\text{max}$. Capture occurs when 
$E_\text{max} > E_R > E_\text{min}$.
It is convenient to write this as
\begin{align}
	\int
\left. dE_R \frac{d\sigma_N}{d E_R}  \right|_{\text{capture}}
&=
\int_{E_\text{min}}^{E_\text{max}} 
dE_R \frac{d\sigma_N}{d E_R}  
\Theta(\Delta E)
&
\Delta E &= E_\text{max} - E_\text{min} \ .
\label{eq:dsig:dEr:capture}
\end{align}
One may then substitute the results of Eqs.~(\ref{eq:w3:f:dw}, \ref{eq:dsig:dEr}, \ref{eq:dsig:dEr:capture}) into Eq.~(\ref{eq:Ccap:N:general}). In Ref.~\cite{Feng:2015hja} we showed that the resulting capture rate may be succinctly written as
\begin{align}
	C_\text{cap}
	&= 32\pi^3 \varepsilon^2 \alpha_X \alpha n_X
	\sum_N \frac{Z_N^2}{m_N E_N} \exp\left(\frac{m_{A'}^2}{2m_N E_N}\right)c_\text{cap}^N
	\\
	c_\text{cap}^N
	&=
\int_0^{R_\odot}  dr \, r^2 n_N(r)
\int_0^\infty du \, u f_\odot(u) 
\Theta(\Delta x_N)
\left[
	\frac{e^{-x_N}}{x_N} + \text{Ei}(-x_N)
	\right]^{x_N^\text{min}}_{x_N^\text{max}} \ ,
\label{eq:DM:capture:rate:full}
\end{align}
where we use the substitution variable $x_N$ and exponential integral
function~\cite{abramowitz1964handbook}, 
\begin{align}
	x_N &= \frac{2 m_N E_R + m_{A'}^2}{2m_N E_N}
& \text{Ei}(z) &\equiv -\int_{-z}^\infty dt \frac{e^{-t}}{t} \ .
\label{eq:Ei}
\end{align}

We use the AGSS09 solar composition model to extract the $n_N(r)$~\cite{Serenelli:2009yc, Serenelli:2009ww, SerenelliWeb}. Ref.~\cite{Baratella:2013fya} tabulated the elements that give the largest contributions to dark matter capture: O, Fe, Si, Ne, Mg, He, S, and N. These are given
in decreasing order of importance, but they are all significant, with the nitrogen contribution just a factor of 5 below that of oxygen in the $m_X \gg m_N$ limit. Hydrogen, the most abundant nucleus in the Sun, is a subdominant target, since the capture rate is proportional to $\mu_N^2 m_N Z_N^2$.

\subsection{Dark Matter Annihilation}

Captured dark matter thermalizes in the Sun for $X$--proton
spin-independent scattering cross sections above $10^{-51}$,
$10^{-50}$, and $10^{-47}~\cm^2$ for $m_X = 100~\gev$, 1 TeV, and 10
TeV, respectively~\cite{Peter:2009mm}.  As we will see below, these values are greatly exceeded here.  The dark matter, then, thermalizes and is Boltzmann distributed in a core near the center of the Sun, with number density
\begin{align}
	n_X(r) &= n_0 e^{-{r^2}/{r_X^2}}
	&
	r_X &= \sqrt{\frac{3T_\odot}{2\pi G_N \rho_\odot m_X}}
	\approx 0.03 \, R_\odot \left(\frac{100~\text{GeV}}{m_X}\right)^{1/2}
\ .
\label{eq:annihilation:parameters}
\end{align}
Writing $\Gamma_\text{ann} = \frac 12 \int d^3 \mathbf{x}\, n_X^2(\mathbf{x}) \langle \sigma_\text{ann} v\rangle$ and using the definition for $C_\text{ann}$ in Eq.~(\ref{eq:NX:Cann}) gives
\begin{align}
	C_\text{ann} &= \langle \sigma_\text{ann} v\rangle \left( \frac{G_N m_X \rho_\odot}{3 T_\odot} \right)^{3/2},
\end{align}
where the solar density and temperature are $\rho_\odot = 151\text{ g/cm}^3$ and $T_\odot = 15.5\times 10^{6}\text{ K}$.  

The captured dark matter is extremely cold, with typical velocity
\begin{equation}
v_0 = \sqrt{2 T_{\odot} / m_X}
= 5.1 \times 10^{-5} \sqrt{\tev/m_X} \ .
\end{equation}
The thermally--averaged $XX \to A' A'$ cross section for
annihilation is therefore significantly modified from the tree-level expression given in Eq.~(\ref{eq:annihilation:rate:Born}) to
\begin{align}
	\langle \sigma_{\text{ann}} v \rangle
	&=
	S\; \langle \sigma_{\text{ann}} v \rangle^\text{Born}
	\ , \label{eq:annihilation:rate}
\end{align}
where $S$ is the non-relativistic Sommerfeld enhancement~\cite{Sommerfeld:1931} of the Born approximation annihilation rate. 
An analytic expression for $S$ for the case of $m_{A'} \neq 0$ may be derived by approximating the Yukawa potential with the
Hulth\'en potential~\cite{Cassel:2009wt,Slatyer:2009vg,Feng:2010zp}, giving an enhancement of $S$-wave processes of
\begin{equation}
	S_s = \frac{\pi}{a} \frac{\sinh(2\pi a c)}{\cosh(2\pi a c) -
          \cos(2\pi\sqrt{c-a^2c^2})} \quad
 	\stackrel{c\gg 1}{\longrightarrow} 
	\quad
	\frac{\pi \, \alpha_X / v} {1 - e^{- \pi \alpha_X / v}} 
          \ ,
\label{eq:sommerfeld:full}
\end{equation}
where $a = v/(2\alpha_X)$ and $c=6\alpha_X m_X/(\pi^2 m_{A'})$. The Sommerfeld enhancement, $S$, is the thermal average of $S_s$, 
\begin{align}
\langle S_S \rangle &= \int \frac{d^3v}{(2\pi v_0^2)^{3/2}} \,
e^{-\frac{1}{2} v^2 / v_0^2} \, S_S  \ .	
\label{eq:sommerfeld:avg}
\end{align}
The general form of $S_s$ on the left-hand side of Eq.~(\ref{eq:sommerfeld:full}) encodes the effects of resonances generated by the long-range potential. Ref.~\cite{Feng:2015hja} showed that these resonances play a crucial role for dark matter accumulation in smaller bodies such as the Earth which would otherwise not be in thermal equilibrium. In contrast, in the regime of parameter space of interest, the Sun is already in thermal equilibrium so that $\tanh \tau_\odot/\tau \approx 1$ in Eq.~(\ref{eq:NX:Cann}) and the effect of the detailed modeling of enhancements to $C_\text{ann}$ is negligible.

\subsection{Dark Matter Self-Capture}

The effect of dark matter self-capture in the Sun, parametrized by $C_\text{self}$ in Eq.~(\ref{eq:rate}), is studied in detail by Zentner in Ref.~\cite{Zentner:2009is}.
$C_\text{self}$ becomes relevant in the regime of very large self-interactions relative to the annihilation rate. One may obtain large self-interactions in the limit of a light mediator since a low-velocity self-interaction enhancement analogous to Sommerfeld enhancement may boost the capture rate; indeed, such a scenario is separately of interest as a proposed solution to small-scale structure anomalies in astrophysics~\cite{Tulin:2013teo}. 

In the dark photon framework discussed here, we find that in the regions of parameter space where a signal is detectable in AMS, the effect of self-capture is negligible. Heuristically, this may be understood as resulting from the fact that, although the self-capture rate is indeed non-perturbatively-enhanced at small velocities, the annihilation rate is Sommerfeld-enhanced even more. This is because the self-scattering occurs with velocities $w \agt v_\odot$, while the annihilation occurs at the much smaller velocities $v_0 \ll v_\odot$ in Eq.~(\ref{eq:sommerfeld:avg}).  As a result, as we will show below, the self-capture contribution to the equilibrium time in Eq.~(\ref{eq:tau}) may be safely ignored. 

For completeness, however, we demonstrate how recent self-interacting dark matter results are applied to self-capture.
Following Ref.~\cite{Tulin:2013teo}, the relevant cross section for self-scattering is the viscosity cross section, $d\sigma_V/d\Omega = \sin^2\theta\, d\sigma/d\Omega$, which regulates forward and backward scattering divergences that do not affect the dark matter phase space evolution. For distinguishable particles, one may approximate this with the transfer cross section, $d\sigma_T/d\Omega = (1-\cos\theta) d\sigma/d\Omega$. The transfer cross section only regulates the forward divergence, but there is an extensive literature on this cross section in the classical limit ($m_X w/m_{A'} \gg 1$) from the plasma physics literature~\cite{PhysRevLett.90.225002,Khrapak:2014xqa}, which may be applied to the present case~\cite{Feng:2009hw}, yielding
\begin{align}
	\sigma_T \simeq \frac{\pi}{m_{A'}^2} \left\{
	\begin{array}{ll}
		4 \beta^2 \ln(1+\beta^{-1})
		&
		\qquad
		\text{if }\beta \alt 10^{-1}
		\\
		4 \beta^2 \left(1+1.5 \beta^{1.65}\right)^{-1}
		&
		\qquad
		\text{if } 10^{-1} \alt \beta \alt 10^{3}
		\\
		\left(\ln \beta + 1 - \frac{1}{2\ln \beta}\right)^2
		&
		\qquad
		\text{if } \beta \agt 10^3
	\end{array}
	\right.
	&
	&
	\beta \equiv \frac{2\alpha_X m_{A'}}{m_X w^2} \ ,
	\label{eq:transfer:xsec:classical}
\end{align}
Over most of the regime for solar dark matter self-capture, $C_\text{self}\sim\sigma_T \sim \alpha_X^2$. Parametrically, 
\begin{align}
	C_\text{cap}C_\text{ann} &\sim \varepsilon^2 \alpha \alpha_X^3 S
	&
	C_\text{self}^2 & \sim \alpha_X^4 \ ,
\end{align}
where $S\sim \alpha_X$. Since both terms scale as $\alpha_X^4$, one cannot tune the dark sector coupling to suppress the ordinary capture rate relative to the self-capture rate. Thus self-capture can only become a dominant effect in the small $\varepsilon$ regime. 
We show below that this only occurs for $\varepsilon$ so small that the Sun is not in equilibrium. In the extreme case, when $\varepsilon$ is so small that $C_\text{cap}C_\text{ann} \ll C_\text{self}^2$, then $\tau \approx 2/C_\text{self}$ in \eqref{eq:tau} is typically much larger than the age of the Sun, $\tau_\odot$, and the annihilation rate is suppressed.

Beyond the classical regime, \eqref{eq:transfer:xsec:classical} must be modified. In the so-called resonant regime where 
\begin{align}
	\frac{\alpha_X m_X}{m_{A'}} & \agt 1
	&
	\text{and}&
	&
	\frac{m_X v}{m_{A'}} & \alt 1 \ ,
\end{align}
one may approximate the Yukawa potential between the dark matter particles with the Hulth\'en potential which may be solved analytically for $S$-wave scattering. In the regime of $m_{A'}/m_X \alt v$, however, higher partial waves are required and one must perform a full numerical integration of the Schr\"odinger equation. A detailed investigation of this limit is beyond the scope of this study.
	
\subsection{Equilibrium Time}

\begin{figure}[h]
\includegraphics[width=0.41\linewidth]{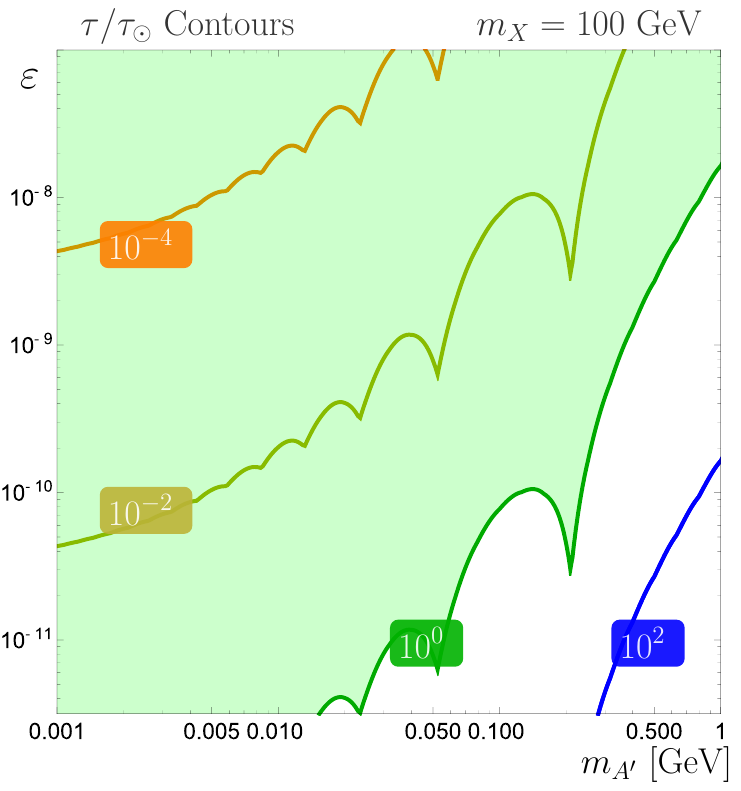} \qquad
\includegraphics[width=0.41\linewidth]{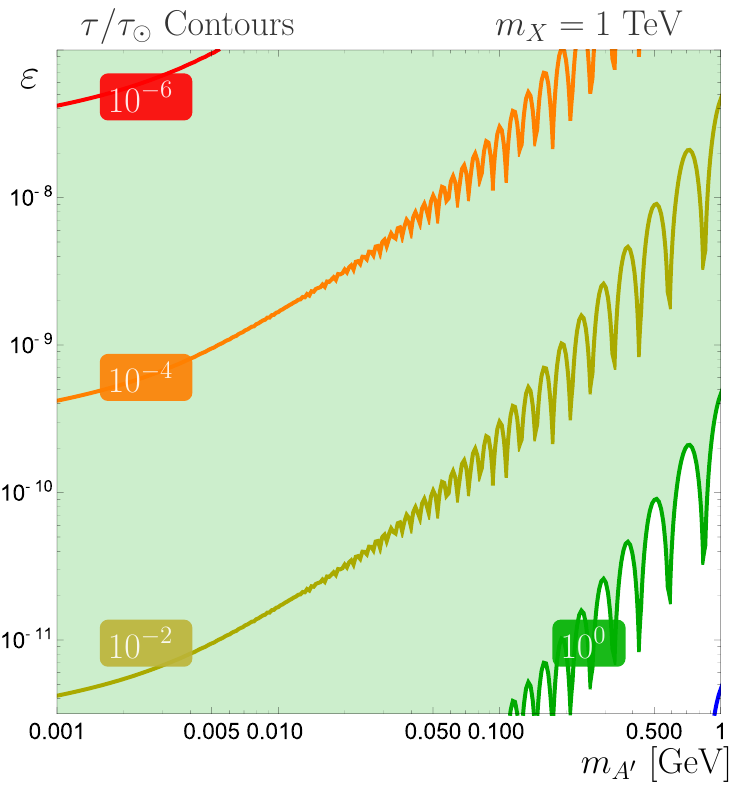} \\
\includegraphics[width=0.41\linewidth]{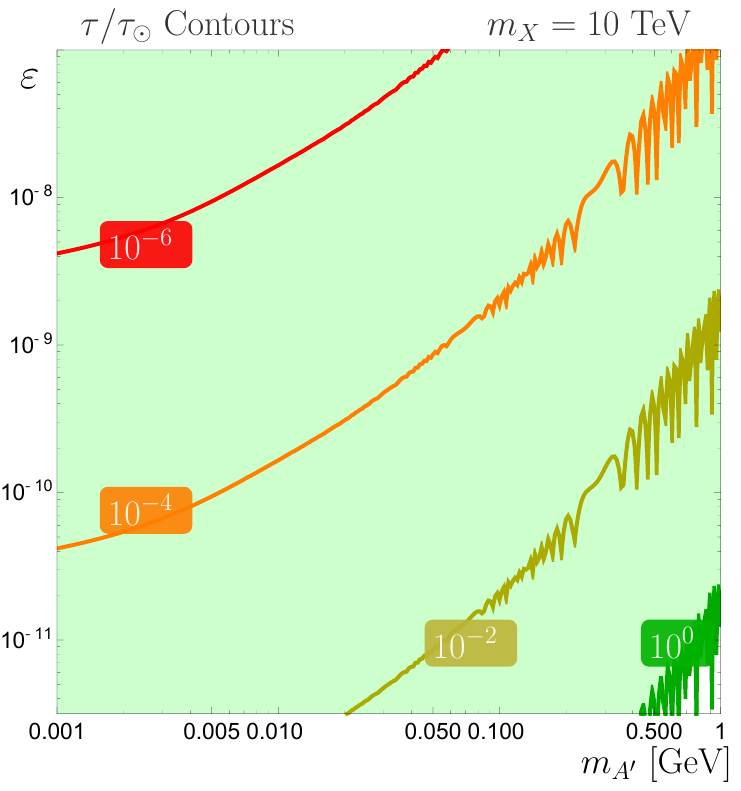} \qquad
\includegraphics[width=0.41\linewidth]{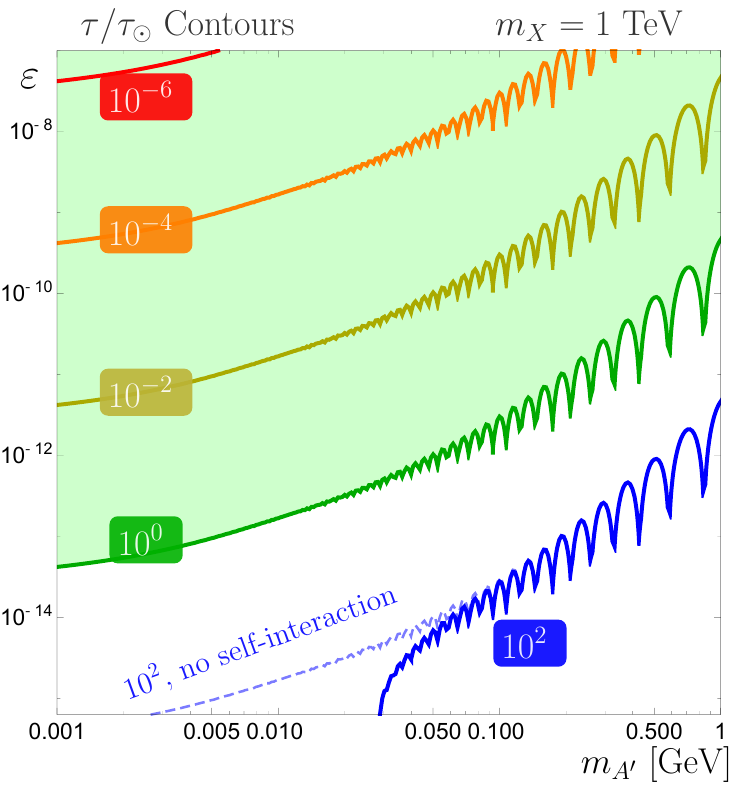} 
\vspace*{-0.1in}
\caption{
Contours of constant $\tau / \tau_\odot$, the equilibrium timescale in units of the Sun's age, in the $(m_{A'}, \varepsilon)$ plane for  $m_X = 100~\gev$  (top left), $1~\tev$ (top right), and $10~\tev$ (bottom left). The dark sector fine-structure constant $\alpha_X$ is set by requiring $\Omega_X \simeq 0.23$. In the green shaded regions, $\tau_\odot < \tau$ and the Sun's dark matter population has reached equilibrium.  \textbf{Bottom--Right}: contours for $m_X=1~\tev$, as in the top right, but extending to very low $\varepsilon$.  The dashed line shows the case where self-capture has been ignored.  The effect of self-capture becomes relevant only for very low $\varepsilon$, where equilibrium times are large and the annihilation signal is highly suppressed.}
  \label{fig:tausun}
\vspace*{-0.1in}
\end{figure}

Fig.~\ref{fig:tausun} presents results for the equilibrium time, $\tau$, defined in Eq.~(\ref{eq:tau}).
The region for which $\tau$ is less than the age of the Sun, $\tau_\odot$, is shaded in green. 
The contours in Fig.~\ref{fig:tausun} reflect the Sommerfeld resonances from Eq.~(\ref{eq:sommerfeld:full}). Unlike the case of the Earth studied in Ref.~\cite{Feng:2015hja}, these resonances do not play a major role since, in the region probed by AMS, $\tanh^2 \tau_\odot/\tau \approx 1$ and the annihilation rate in Eq.~(\ref{eq:NX:Cann}) is not affected by further enhancements.
The bottom right plot shows the regime where self-interactions are significant and cause a noticeable deviation from the $C_\text{self} = 0$ limit. As noted above, this only occurs in the region where the Sun is not yet in equilibrium so that the dark matter annihilation rate is suppressed.

\section{Positron Signal and Background at AMS} 
\label{eq:signal:and:background}

\begin{figure}[tb]
\includegraphics[width=.99\linewidth]{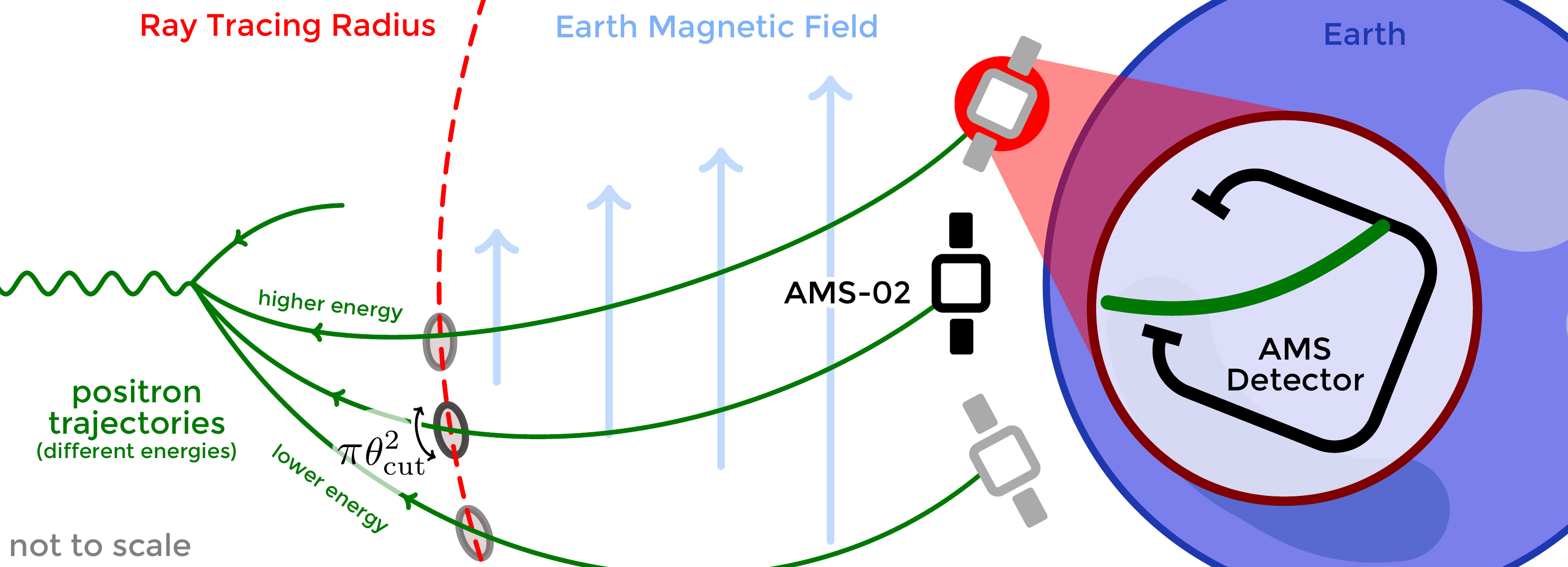}
\vspace*{-0.1in}
  \caption{Schematic depiction of positron trajectories bending in the Earth's magnetic field. For each positron energy, one considers a solid angle $\pi \theta^2_\text{cut}$ given by Eq.~(\ref{eq:theta:cut}). Since the Earth's magnetic field is well known this mapping is well defined. The inset shows the origin of the angular dependence implicit in the Sun exposure in Eq.~(\ref{eq:exposure}).
}
  \label{fig:picture:earth:field}
\vspace*{-0.1in}
\end{figure}

The dark photons produced by dark matter annihilation in the Sun decay to all kinematically-accessible charged SM particles, leading to a variety of possible signals ($e^+ e^-$, $\mu^+ \mu^-$, $\pi^+ \pi^-$, etc.) that can be detected in a number of experiments. We consider the $e^+e^-$ signal for dark photons with $m_{A'} > 2m_e$. We specifically focus only on positrons, since the $e^+$ and $e^-$ signals have identical properties and the positron background is smaller~\cite{Aguilar:2014mma}, and we consider the AMS-02 experiment on the International Space Station (ISS), which is optimal for positron detection.

The positron signal and background are very different: the signal has a hard spectrum and points back to the Sun, while the astrophysical
background drops rapidly with energy and is effectively isotropic.  In principle, it is therefore easy to isolate the signal by considering very energetic positrons that point back to the Sun.  In practice, however, the signal is greatly complicated by the magnetic fields of the Sun and Earth, which each significantly deflect even TeV positrons.  In the following, we begin by accounting for the Sun's magnetic field, which is not well constrained, and neglecting the Earth's magnetic field, which is relatively well understood.   

Our general strategy is the following: for fixed parameters $m_X$, $m_{A'}$, and $\varepsilon$, and a given experimental live time $T$, we consider only positrons with energies above $E_{\text{cut}}$ that point back to the Sun within an angle $\theta_{\text{cut}}$.  For a particular choice of $E_{\text{cut}}$, we choose $\theta_{\text{cut}} ( E_{\text{cut}})$ so that the number of background positrons is $N_B=1$. We then determine the number of signal positrons, $N_S$, that pass these cuts, given a model for the Sun's magnetic field.  We then determine the optimal value of the energy cut, $E_{\text{cut}}^{\text{opt}}$, which maximizes $N_S$, and we use these maximal values of $N_S$ to determine the reach of AMS. 

This procedure neglects the Earth's magnetic field.  Since this magnetic field is well mapped, we assume that its effect on the signal may be de-convoluted so that positrons can be ray-traced back to a distance of several $R_{\oplus}$ from the Earth, where the Earth's magnetic field is negligible. It is at this position that the solid angle of size $\pi \theta_\text{cut}^2$ should be defined. This is shown schematically in Fig.~\ref{fig:picture:earth:field}. 

In the remainder of this section we discuss the number of background events $N_B$, the bending of positrons in the solar magnetic field, the number of signal events $N_S$, and the optimization of $N_S$.  

\subsection{Number of Background Events: Energy and Angular Cuts}
\label{sec:background}

We define the signal to be positrons with energies above $E_{\text{cut}}$ that point back to the Sun within an angle $\theta_{\text{cut}}$. Together, these parameters control the number of background positrons.
The background isotropic positron flux has been precisely measured by
AMS~\cite{Aguilar:2014mma} to be
\begin{align}
\frac{d\Phi}{dE} &\approx \frac{1.5 \times 10^{-9}}{\gev~\cm^2~\sr~\s}
\left( \frac{E}{100~\gev} \right)^{-2.8} \ .
\label{eq:dPhidE}
\end{align}
The number of background events in the signal region
is, then,
\begin{align}
	N_B(E_\text{cut},\theta_\text{cut}) = \xi_\odot \, \Omega_{\odot}(\theta_\text{cut})
	\int_{E_\text{cut}}^\infty \frac{d\Phi}{dE} dE
	\ ,
\end{align}
where, for small $\theta_{\text{cut}}$, 
\begin{equation}
\Omega_\odot(\theta_\text{cut}) 
= \pi \, \theta_{\text{cut}}^2 ~\sr
\simeq 
	9.6 \times 10^{-4}~\sr
	\left(\frac{\theta_\text{cut}}{1^\circ}\right)^2
\end{equation}
is the solid angle subtended by $\theta < \theta_\text{cut}$, and
$\xi_\odot $ is the exposure of AMS to the Sun, a function of positron energy, the ISS's orbit, and AMS's fixed orientation on the ISS.  For positron energies above 50 GeV, a detailed calculation finds that in 924 days of livetime, AMS's exposure to the Sun was $\xi_{\odot} \simeq 1.6 \times 10^5~\m^2~\s$~\cite{Machate}.  Assuming uniform operating conditions, then,
\begin{equation}
	\xi_\odot = 
	6.3 \times 10^4~\m^2~\s \,
	\frac{T}{\yr}
	\simeq 20~\cm^2 \ T \ ,
		\label{eq:exposure}
\end{equation}
where $T$ is the AMS livetime, that is, its total time in orbit.  
The ``effective area'' 20 cm$^2$ is much smaller than the geometric size of the detector 
due, in part, to the fact that the Sun is only in the field of view a small fraction of the time. For comparison, if AMS spent 100\% of its livetime with the sun at the center of its field of view, the exposure would be about 80 times larger~\cite{Machate}.

The resulting number of background events is
\begin{align}
N_B(E_\text{cut},\theta_\text{cut}) 
&= 
0.051
\left(\frac{100~\text{GeV}}{E_{\text{cut}}}\right)^{1.8}
\left(\frac{\theta_{\text{cut}}}{1^\circ}\right)^2
\left(\frac{T}{\yr}\right)
	\ .
\end{align}
Fixing $\theta_\text{cut}$ as a function of $E_\text{cut}$ for a given $T$ by requiring only a single background event, $N_B = 1$, yields
\begin{align}
	\theta_\text{cut}(E_\text{cut})
	&= 
	4.4^\circ 
	\left( \frac{E_\text{cut}}{100~\gev} \right)^{0.9}
	\left( \frac{\yr}{T} \right)^{1/2}
	\ .
	\label{eq:theta:cut}
\end{align}

\subsection{Bending of Signal Positrons by the Solar Magnetic Field} \label{sec:deflection}

Before quantifying the number of signal events, let us examine the bending of a signal positron by the solar magnetic field.
In the absence of magnetic fields between the Sun and the Earth, positrons from solar dark photon decays would point back to within a degree (for $m_X > 100$~GeV) of the center of the Sun where the dark matter is concentrated within a core of radius $r_X$ from Eq.~(\ref{eq:annihilation:parameters}). 
The AMS electromagnetic calorimeter's angular resolution is parametrized by $\Delta \theta_{68} \simeq \sqrt{5.8^{\circ 2} / (E~\text{in GeV}) + 0.23^{\circ
    2}}$~\cite{Gallucci:2015fma}; the angular resolution from the tracker is even better~\cite{Machate}.  For the positron energies we will consider, the experimental angular resolution is therefore less than a degree and is negligible.  
    
The signal, however, is smeared out by the solar magnetic field, which bends charged particles as they travel to the Earth. 
Because of the solar wind, the magnetic field of the Sun differs from a dipole and varies with the 11-year solar cycle. As an approximation, we use the Parker model for the heliospheric magnetic field, which has radial and azimuthal components in heliocentric coordinates~\cite{1958ApJ...128..664P}; see Refs.~\cite{lrsp-2013-3, balogh2007heliosphere} for reviews. Since the positrons propagate in the radial direction, it is sufficient to model the azimuthal part of the magnetic field,
\begin{align}
	B_\phi = \left(\frac{3.3~\text{nT}}{\sqrt{2}}\right) \frac{\text{au}}{r} \ ,
\end{align}
where we have used the facts that at $R=\text{au}$, $|\vec B| = 3.3~\text{nT}$ and the radial and azimuthal components of the field are equal in magnitude. We ignore a subleading $r^{-2}$ piece in $B_\phi$ which is suppressed by a factor of $R_\odot = 0.005\text{ au}$. This model was invoked in Ref.~\cite{Roberts:2010yh} to explain the PAMELA positron excess as the result of increased activity during the solar cycle.
With this magnetic field, the bending angle of a positron of energy $E$ produced at a dark photon decay position $r_d$ from the Sun is 
\begin{align}
	\theta_\text{bend}(r_d,E) &= 
	8.9^\circ \left(\frac{\text{TeV}}{E}\right)
	\int_{r_d}^\text{au} \frac{B_\phi(r')\,dr'}{\text{au} \, (3.3~\text{nT})}
	= 6.3^\circ \left(\frac{\text{TeV}}{E}\right) \ln \frac{\text{au}}{r_d}
	\ .
	\label{eq:theta:bend}
\end{align}

\subsection{Number of Signal Events}
\label{sec:signal}

The total number of signal events $N_S$ is
\begin{align}
	N_S = N_S^0 \, 
	\text{Br}(A'\to e^+e^-) \, 
	\Pdet \ ,
	\label{eq:NS}
\end{align}
where
\begin{align}
N_S^0
= 2 \Gamma_\text{ann} \, \frac{\xi_\odot}{4 \pi (1~\text{au})^2}
\label{eq:NS0}
\end{align}
is the number of dark photons produced when the Sun is in AMS's field of view,  $\text{Br}(A'\to e^+e^-)$ is the probability that a dark photon decays to a positron, and $\Pdet$ is the probability that such a positron is detected within the signal region by AMS.  In \eqref{eq:NS0}, the factor of 2 accounts for the two dark photons produced per dark matter annihilation, and $\xi_{\odot}$ is the exposure defined in \eqref{eq:exposure}.  $N_S^0$ and $\text{Br}(A'\to e^+e^-)$ are completely determined by the model parameters, while $\Pdet$ depends also on the cut parameters.

\begin{figure}[tb]
	\includegraphics[width=.7\linewidth]{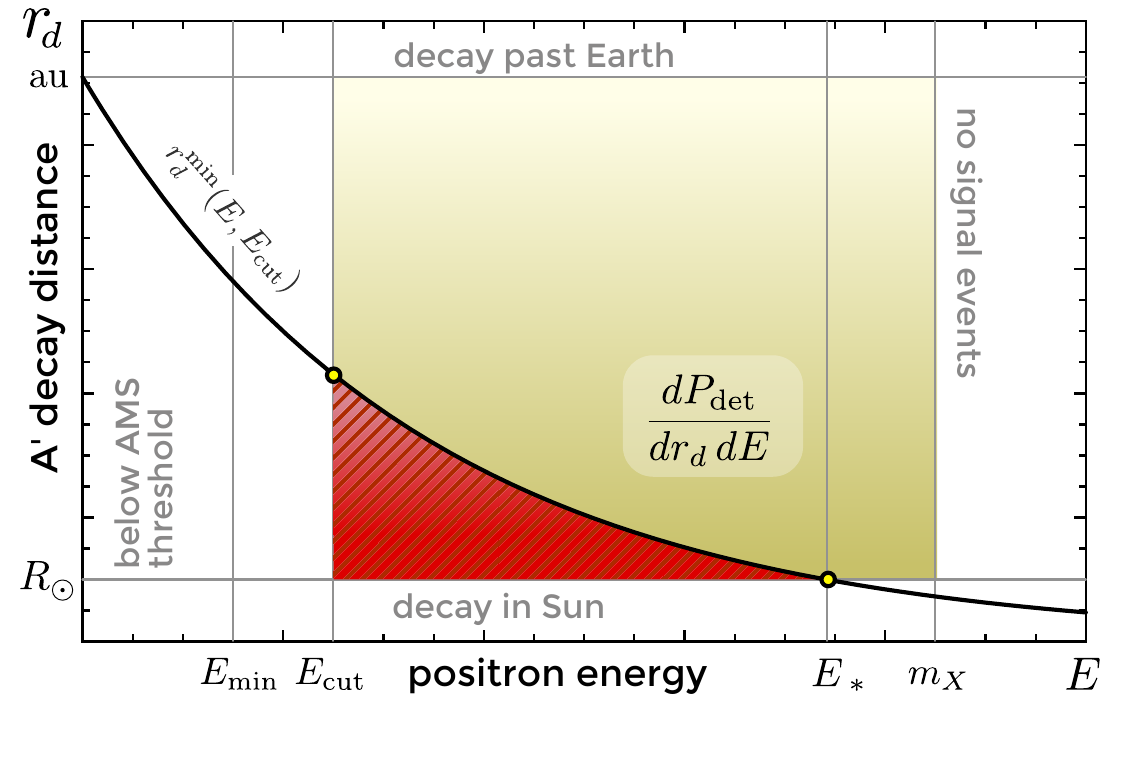}
	\vspace*{-0.1in}
	\caption{Schematic depiction of the signal region of integration, Eq.~(\ref{eq:ROI}), in the plane of $A'$ decay distance $r_d$ and positron energy $E$. The beige shading represents the magnitude of the integrand, Eq.~(\ref{eq:pdf}). 
	We integrate over the box $R_\odot < r_d < \text{au}$, $E_\text{cut} < E < m_X$ and then subtract the integral over the red shaded region bounded by $R_\odot$ and $r_d^\text{min}$. 
	}
	\label{fig:ROIJordan}
	\vspace*{-0.1in}
\end{figure}

We now determine the detection probability $\Pdet$.  For a positron to be detected in the AMS signal region, (1) it must be created by a dark photon that decays after traveling a distance between $R_{\odot}$ and 1 au, and (2) it must not be deflected out of the signal region by the solar magnetic field. Letting $r_d$ be the distance a dark photon travels before it decays.  condition (2) implies 
\begin{align}
	\theta_\text{bend}(r_d,E) \leq \theta_\text{cut}(E_\text{cut})
	\ ,
\end{align}
or, given Eqs.~(\ref{eq:theta:cut}) and (\ref{eq:theta:bend}), 
\begin{align}
	r_d \geq r_d^\text{min}(E,E_\text{cut}) 
	&\equiv
	~\text{au} ~ e^{-E/E_0(E_\text{cut})} 
	\ ,
	\label{eq:rmin}
\end{align}
where
\begin{align}
E_0(E_\text{cut}) &\equiv 1.5~\text{TeV} \left(\frac{100 ~\text{GeV}}{E_\text{cut}}\right)^{0.9} \left(\frac{T}{\yr} \right)^{1/2} \ .
\end{align}
Positrons that do not satisfy Eq.~(\ref{eq:rmin}) are produced too far from the Earth and are deflected too much to satisfy the angle cut.
Given the two constraints on $r_d$, the signal region in the space of dark photon decay position $r_d$ and positron energy $E$ is bounded by
\begin{align}
	R_\odot & \leq r_d \leq \text{au} 
	&
	r_d^\text{min}(E,E_\text{cut}) & \leq r_d
	&
	E_\text{min} \leq E_\text{cut} &\leq  E \leq m_X 
	\label{eq:ROI}
	\ ,
\end{align}
where $E_\text{min} = 50~\text{GeV}$ is the minimum positron energy cut from AMS. This region is shown in Fig.~\ref{fig:ROIJordan}. 

The probability density for positrons to be produced at position $r_d$ and energy $E$ is
\begin{align}
	\frac{d \Pdet}{dr_d \, dE}  &=  
	\frac{e^{-r_d/L}}{L} \frac{1}{m_X} \ ,
	 \label{eq:pdf} 
\end{align}
where the decay length, $L$, is defined in Eq.~(\ref{eq:decay:length}), and we have used the fact that for $m_e \ll m_{A'} \ll m_X$, the positron energies are evenly distributed in the range $0\leq E \leq m_X$. Ref.~\cite{Ajello:2011dq} confirms that these positrons do not lose appreciable energy propagating to Earth.

The probability for a positron to be detected in the AMS signal region is, then, 
\begin{align}
	\Pdet
	&= 
	\int \frac{d \Pdet}{dr_d \, dE} \, dr_d \, dE 
	=
	\int e^{-r_d/L} \, \frac{dr_d}{L} \frac{dE}{m_X} \, 
	\equiv 
	\Pdet^0 -  \Pdet^B\, \Theta(E_* - E_\text{cut}) \ ,
	\label{eq:acceptance}
\end{align}
where the region of integration is defined by Eq.~(\ref{eq:ROI}).
$\Pdet^0$ and $\Pdet^B$ are defined to be the integral over the box and the red region, respectively, in Fig.~\ref{fig:ROIJordan}.
$\Pdet^0$ is the probability, in the absence of magnetic fields, that a dark photon will decay after traveling a distance between $R_\odot$ and $1~\text{au}$ to produce a positron with energy greater than $E_\text{cut}$. $\Pdet^B$ is the correction to this na\"ive probability caused by the angular cuts to account for the solar magnetic field. 
$E_*$ is defined to be the energy for which $r_d^\text{min}(E_*,E_\text{cut}) = R_\odot$. Above this energy the condition Eq.~(\ref{eq:rmin}) is trivial since dark photons must decay beyond $R_\odot$ or else their decay products are caught in the Sun.
The upper limit of the $dE$ integral in $\Pdet^B$  is
\begin{align}
	E_\times &\equiv \text{min}\left(E_*, m_X\right)
	&
	&\text{ where } 
	&
	E_* &= E_0 \log \frac{\text{au}}{R_\odot} \ .
\end{align}
This definition of $E_\times$ is necessary since $E_* > m_X$ for sufficiently small $E_\text{cut}$.
For the Parker model of the solar magnetic field, the integrals can be evaluated exactly:
\begin{align}
	\Pdet^0 &=
	\frac{m_X - E_\text{cut}}{m_X}\left(e^{-\frac{R_\odot}{L}} - e^{-\frac{\text{au}}{L}}\right)
	\label{eq:A0}
	\\
	\Pdet^B &=
	\frac{E_0}{m_X}\left[
	\text{Ei}\left(-\frac{\text{au}}{L} e^{-\frac{E_\times}{E_0}}\right) - \text{Ei}\left(-\frac{\text{au}}{L} e^{-\frac{E_\text{cut}}{E_0}}\right)
	\right] + \frac{E_\times-E_\text{cut}}{m_X}e^{-\frac{R_\odot}{L}} \ ,
\label{eq:AB}
\end{align}
where the Ei function is defined in Eq.~(\ref{eq:Ei}).
 
\begin{figure}[tb] 
	\hspace*{-.5cm}
	\includegraphics[width=0.45\linewidth]{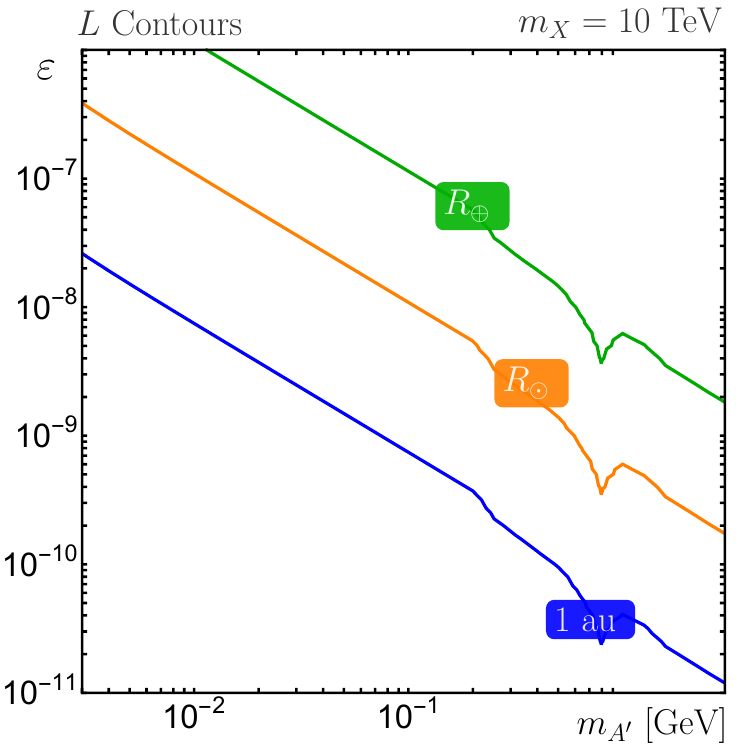} \qquad
	\includegraphics[width=0.45\linewidth]{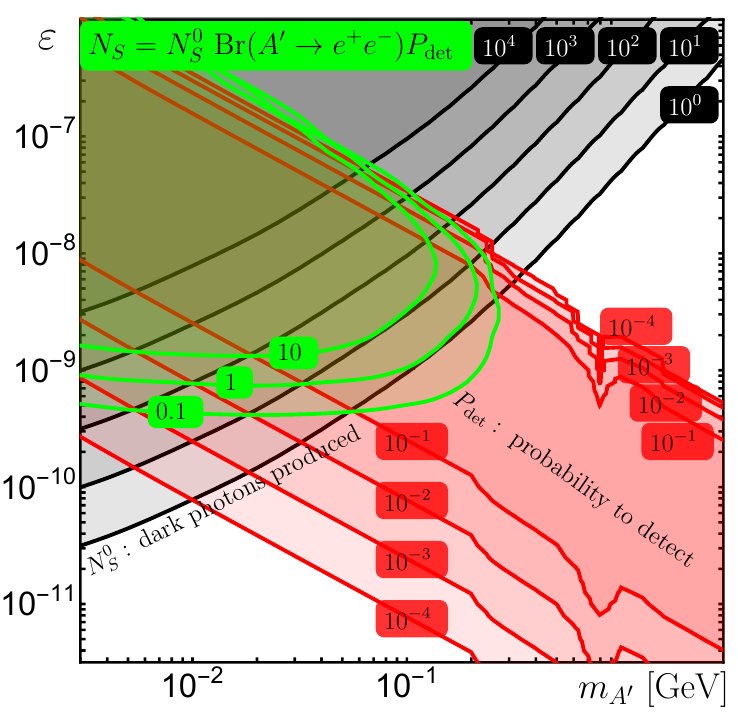} \qquad 
	\vspace*{-0.1in} 
	\caption{
		\textbf{Left:} Contours of fixed decay length $L = R_\oplus \simeq 6400~\km$, $R_{\odot} \simeq 7.0 \times 10^6~\m$, and 1 au in the $(m_{A'}, \varepsilon)$ plane for $m_X = 10~\text{TeV}$. The dip at 775~MeV comes from resonant $A'$ decays via $\rho$ mesons mixing. The decay lengths shape the probability contours through Eq.~(\ref{eq:A0:approx}).
		\textbf{Right:} Factors determining the signal reach for dark sunshine searches at AMS for livetime $T=$ 3 years and $m_X = 10~\tev$. \textsc{Black}:  $N_S^0$, the number of dark photons produced when the Sun is in AMS's field of view. \textsc{Red}: $\Pdet^\text{max}$, the optimal positron detection probability at each point in the $(m_{A'}, \varepsilon)$ plane. \textsc{Green}: $N_S$, signal region reach. In most of this plane, $\text{Br}(A'\to e^+e^-) =1$, and so the green contours are products of the red and black contours.  For example, the $N_S$ = 1 contour passes through the intersection of the $N_S^0 = 10$ and $\Pdet^\text{max} = 0.1$ contours.}
	\label{fig:acceptancediscussion}
	\vspace*{-0.1in}
\end{figure}

The difference of exponentials in Eq.~(\ref{eq:A0}) determines the shape of the region of dark photon parameter space that can be reached. When $L \ll R_\odot$ this term drops rapidly because few dark photons decay outside the Sun. When $L \gg \text{au}$, one may expand the exponentials so that
\begin{align}
	\Pdet^0 \approx 
	\frac{m_X - E_\text{cut}}{m_X} \, \frac{\text{au}}{L}
	\propto \varepsilon^2 m_{A'}^2 \ . 
	\label{eq:A0:approx}
\end{align}
This is illustrated in Fig.~\ref{fig:acceptancediscussion}, which shows contours of constant decay length $L$ and how these shape the $N_S^0$ and $N_S$ reach. Values of $E_\text{cut}$ are chosen for each choice of $(m_{A'},\varepsilon)$ to optimize $\Pdet$.
Decreasing the dark matter mass $m_X$ produces lower energy positrons which are subsequently deflected more by the magnetic fields so that the probability decreases. For example, at $m_X = 100~\text{GeV}$ the maximum probability is reduced by two orders of magnitude relative to $m_X = \text{TeV}$, significantly reducing the reach of AMS.

\subsection{Optimizing the Signal}
\label{sec:optimizing}

\begin{figure}[t]
	\includegraphics[width=.65\linewidth]{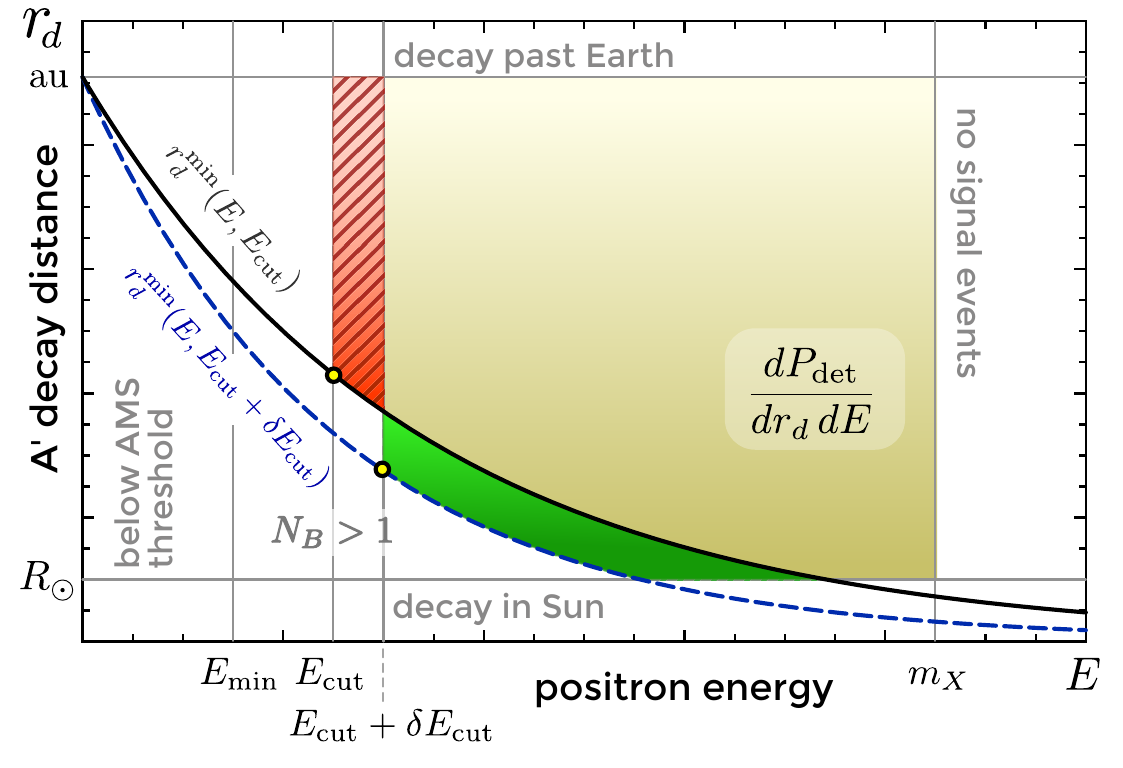}
	\vspace*{-0.1in}
	\caption{Schematic description of the signal region in the plane of decay distance $r_d$ and positron energy $E$. As one varies $E_\text{cut}$, the $r_d^\text{min}$ line shifts downward while the lower limit of the $dE$ integration shifts upward. The optimal $E_\text{cut}$ is then when the integral over the red and green regions are equivalent. The shading represents the magnitude of the integrand, Eq.~(\ref{eq:pdf}).}
	\label{fig:ROI}
	\vspace*{-0.1in}
\end{figure}

\begin{figure}[h] 
	\vspace*{.5cm}
	\includegraphics[width=0.45\linewidth]{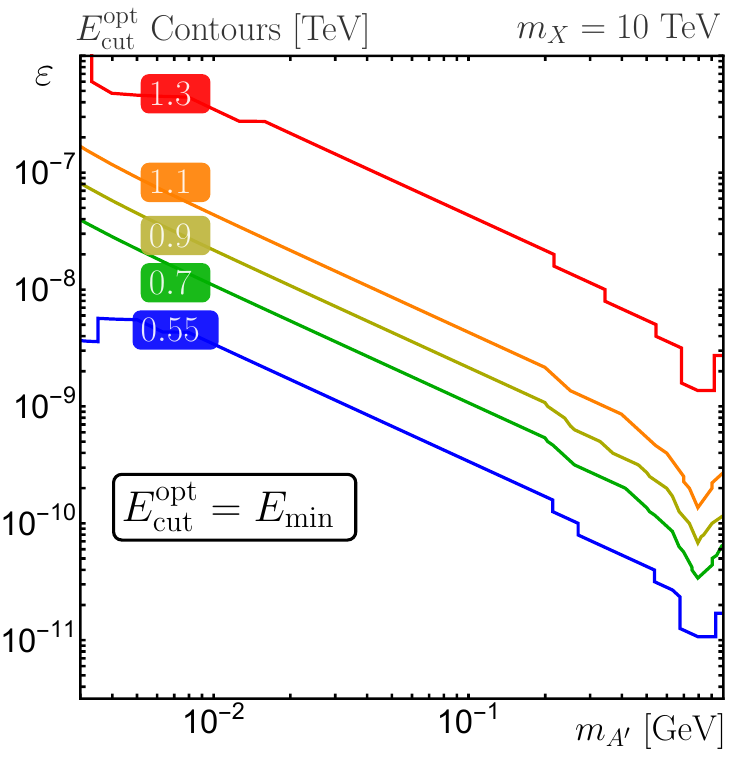} \qquad
	\includegraphics[width=0.45\linewidth]{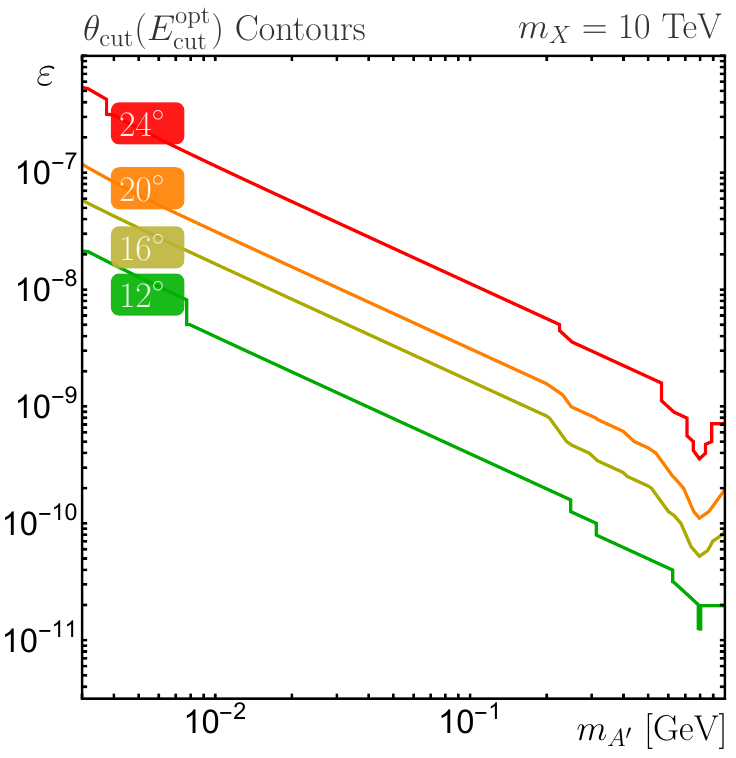}
	\vspace*{-0.1in} 
	\caption{
		\textbf{Left:} Contours of $E_\text{cut}^{\text{opt}}$, the value of $E_{\text{cut}}$ that maximizes the probability $\Pdet$ in the $(m_{A'}, \varepsilon)$ plane for $m_X = 10~\text{TeV}$. In the region below lowest plotted contour, $550 ~\text{GeV}$, $E_\text{cut}^{\text{opt}} > E_*$ so that the $\Pdet^B$ term in Eq.~(\ref{eq:acceptance}) vanishes---the solar magnetic field does not affect the choice of cuts---and $E_\text{cut}^\text{opt} = E_\text{min}$. 
		\textbf{Right:} Contours of the corresponding values of $\theta_\text{cut}$ from Eq.~(\ref{eq:theta:cut}). 
	}
	\label{fig:bestcuts10TeV}
	\vspace*{-0.1in}
\end{figure}

Throughout this study we choose $E_\text{cut}$ to optimize the signal probability $\Pdet$ while fixing the number of background events, $N_B = 1$. Because the probability is a concave function of $E_\text{cut}$, the choice of $E_\text{cut}$ as a function of $m_X, m_{A'}$, and $\varepsilon$ is found by solving $d\Pdet/dE_\text{cut} = 0$, where
\begin{align}
	\frac{d\Pdet}{dE_\text{cut}}
	&=
	-\frac{1}{m_X} \left(e^{-\frac{R_\odot}{L}} - e^{-\frac{\text{au}}{L}}\right)
	+ \frac{1}{m_X} \Theta(E_* - E_\text{cut}) 
	\bigg[
	\mathcal F_1
	\Theta(m_X - E_*) 
	+
	\mathcal F_2
	\bigg] 
\\
\mathcal F_1 & = \phantom{+} \frac{0.9 E_0}{E_\text{cut}} 
\left[
	e^{-\frac{R_\odot}{L}} - \exp\left(-\frac{\text{au}}{L} e^{-\frac{E_\times}{E_0}}\right)
	\right] 
	\log\frac{\text{au}}{R_\odot}
\\
\mathcal F_2 & = \phantom{+} e^{-\frac{R_\odot}{L}} - \exp\left(-\frac{\text{au}}{L}e^{-\frac{E_\text{cut}}{E_0}}\right)
\ .
\end{align}
This equation is solved numerically to give the choice $E_\text{cut}^{\text{opt}}$ that optimizes $\Pdet$. To clarify the nature of this optimization, we show the effect of varying $E_\text{cut}$ in Fig.~\ref{fig:ROI}.
Figure~\ref{fig:bestcuts10TeV} shows a set of representative $E_\text{cut}^{\text{opt}}$ contours in the $(m_{A'}, \varepsilon)$ plane for $m_X = 10~\text{TeV}$. 

\section{Results: AMS Reach} 

\begin{figure}[h] 
	\hspace*{-.5cm} 
	\includegraphics[width=0.41\linewidth]{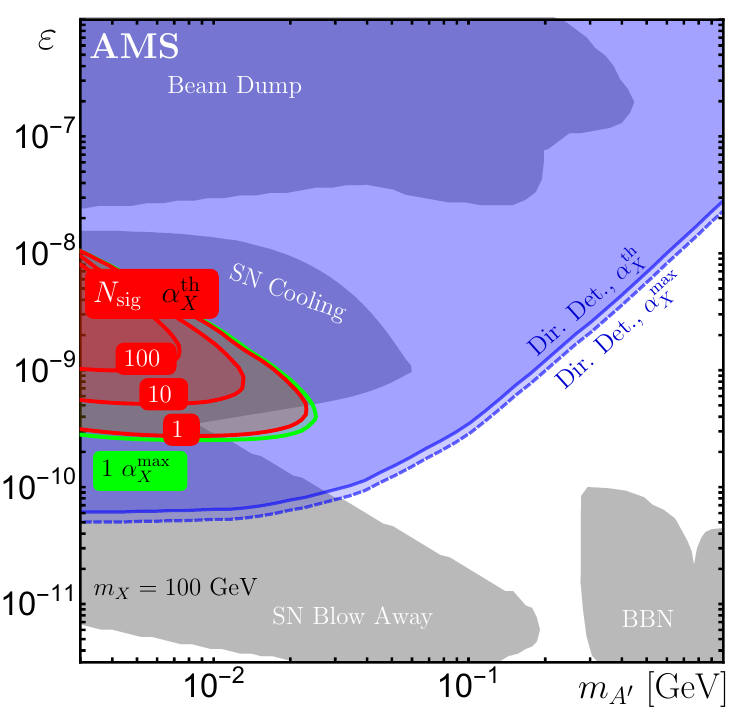} \qquad
	\includegraphics[width=0.41\linewidth]{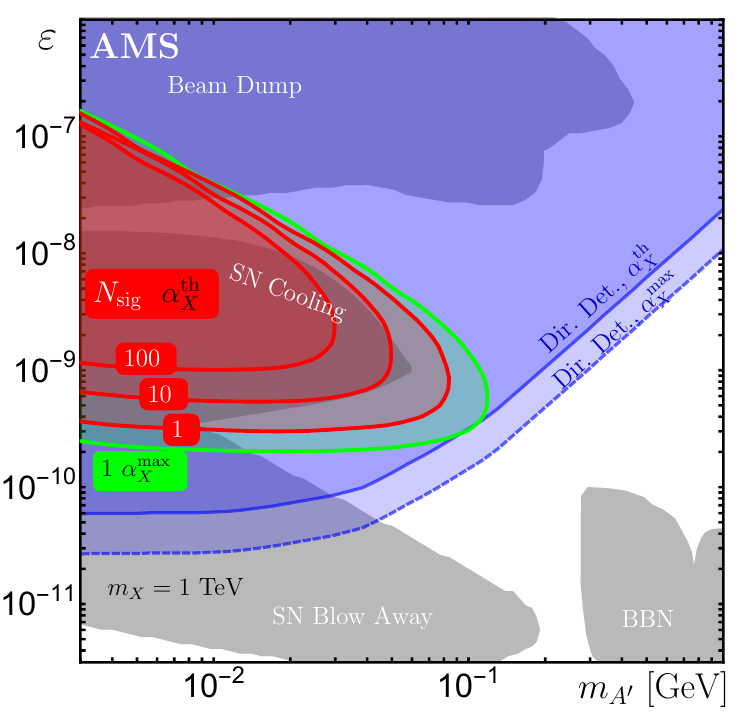} \\
	\includegraphics[width=0.41\linewidth]{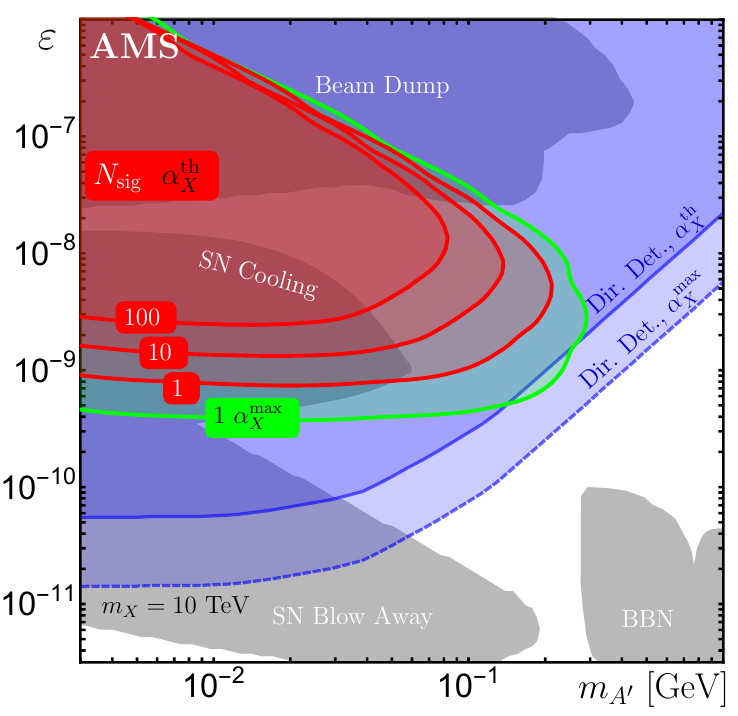} \qquad
	\includegraphics[width=0.41\linewidth]{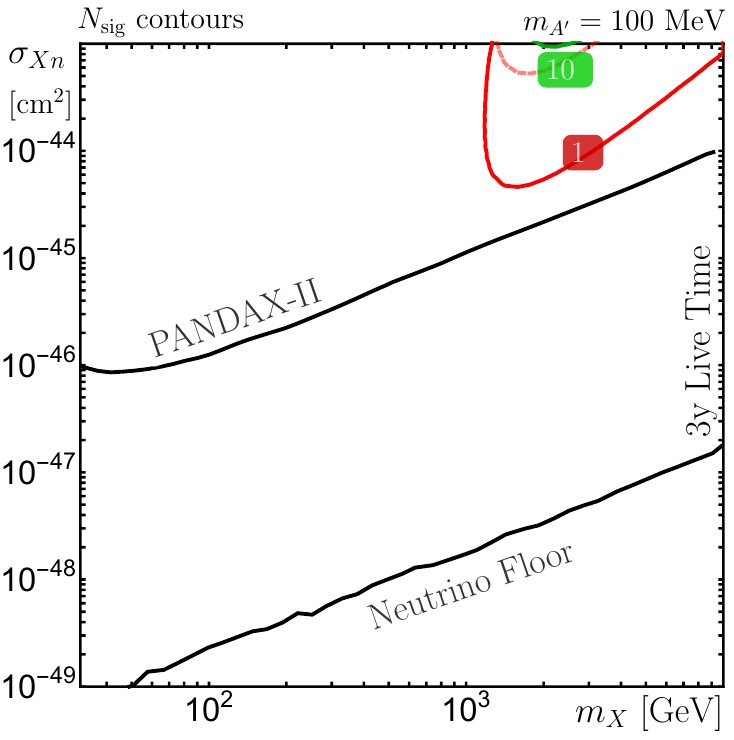} 
	\vspace*{-0.1in} 
	\caption{ \textbf{Top} and \textbf{Bottom--Left}:
		\textsc{Red:} Number of AMS signal events $N_S$ for $m_X = 100~\gev , 1~\tev , 10~\tev$, $N_B = 1$ background event, and livetime $T = 3$ years in the $(m_{A'}, \varepsilon)$ plane. The dark sector fine-structure constant $\alpha_X$ is set
		by requiring $\Omega_X \simeq 0.23$.  \textsc{Green:} The $N_S = 1$ reach for $\alpha_X = \alpha_X^\text{max}$, the maximal allowed coupling from CMB bounds~\cite{Adams:1998nr,Chen:2003gz,Padmanabhan:2005es}, as written in Eq.~(\ref{eq:alphaX:max}).  \textsc{Blue:} Current bounds from  direct detection~\cite{DelNobile:2015uua,Cui:2017nnn}.  \textsc{Gray:} Regions probed by other dark photon searches discussed in \secref{bounds}.
		\textbf{Bottom--Right}: Comparison of indirect and direct detection sensitivities
		in the $(m_X, \sigma)$ plane for $m_{A'} = 100~\mev$. \textsc{Red}: $N_S=1$ signal event contours for $\alpha_X = \alpha_X^\text{th}$ (solid) and $\alpha_X^{\text{max}}$ (dashed).  \textsc{Green}: Same, but for $N_S = 10$. The direct detection bounds are from the PANDAX-II experiment~\cite{Cui:2017nnn}; note that in this regime the point-like interaction limit is valid; this is not the case for the low $m_{A'}$ region~\cite{Kaplinghat:2013yxa, An:2014twa, DelNobile:2015uua}.  Also shown is the ``neutrino floor,'' where coherent neutrino scattering affects direct detection experiments~\cite{Billard:2013qya}.
	}
	\label{fig:results}
	\vspace*{-0.1in}
\end{figure}

To provide a rough estimate of AMS's discovery potential, in Fig.~\ref{fig:results} we show results for the number of signal events  that pass the optimized cuts detailed in the previous section. Contours of $N_S$ are given in the $(m_{A'},\varepsilon)$ plane for both thermal ($\alpha_X^\text{th}$) and maximal ($\alpha_X^\text{max}$) dark sector couplings and for the benchmark dark matter masses, $m_X = 100~\gev$, TeV, and 10~TeV. These are the same benchmark masses used in our recent analysis of Earth capture of dark matter~\cite{Feng:2015hja}. 

\begin{figure}[h] 
	\hspace*{-.5cm} 
	\includegraphics[width=0.41\linewidth]{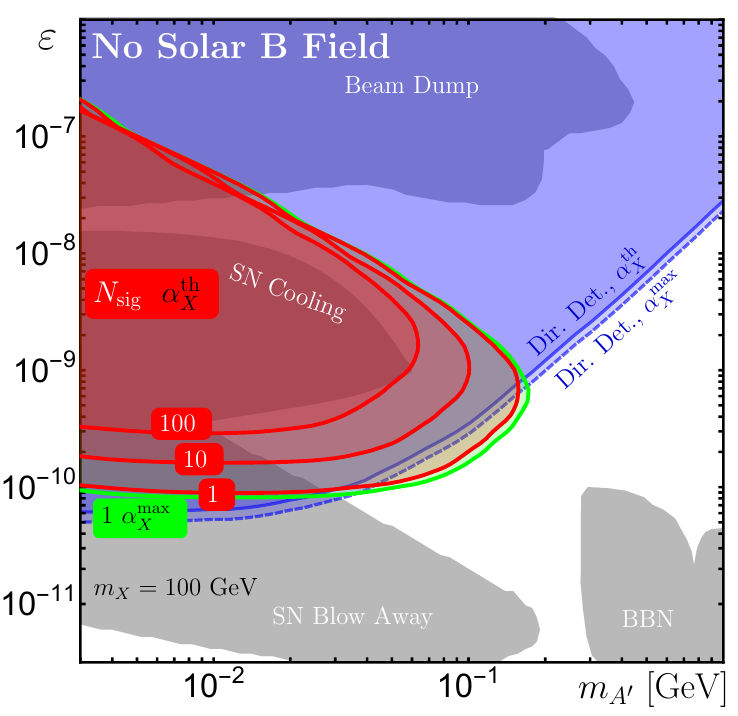} \qquad
	\includegraphics[width=0.41\linewidth]{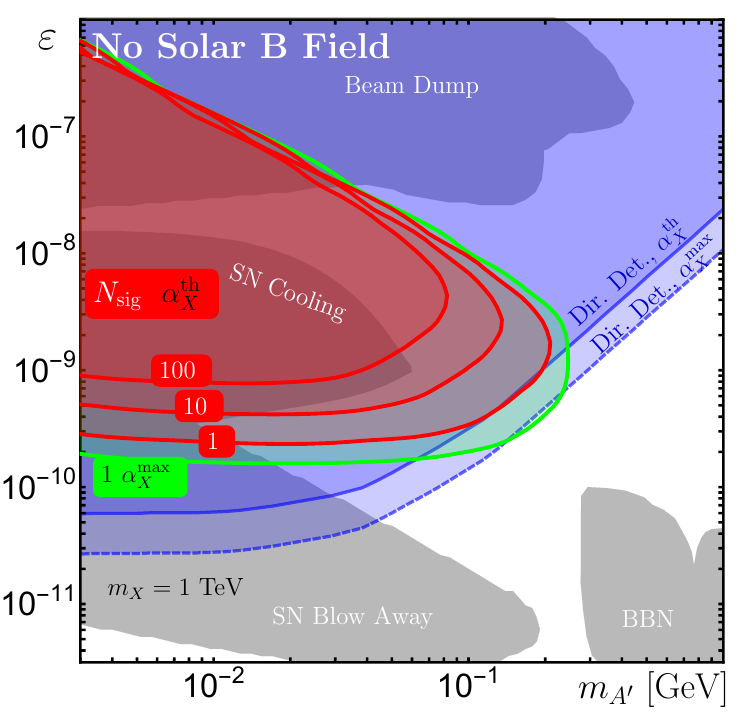} \\
	\includegraphics[width=0.41\linewidth]{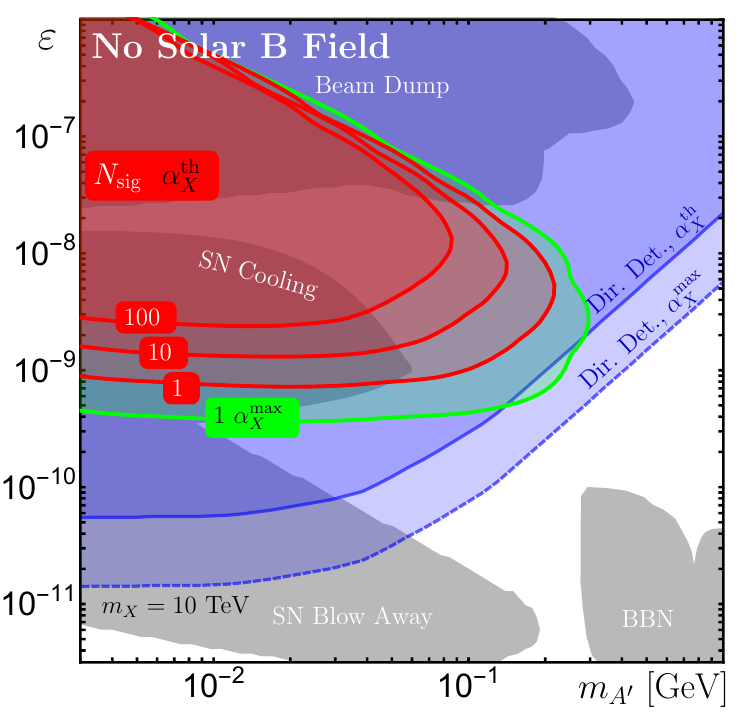} \qquad
	\includegraphics[width=0.41\linewidth]{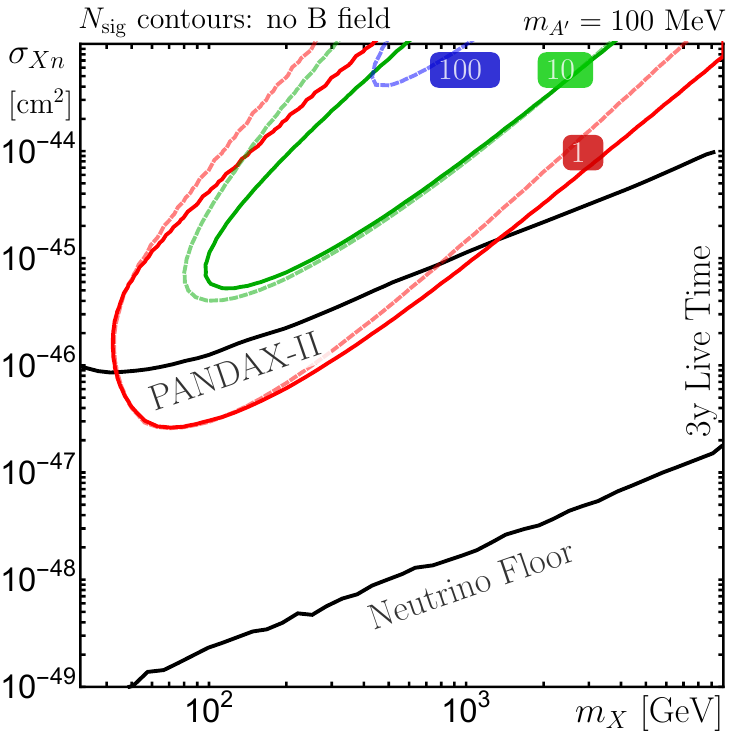} 
	\vspace*{-0.1in} 
	\caption{Same as Fig.~\ref{fig:results} but with no solar $B$ field. Comparing to Fig.~\ref{fig:results}, one sees that a large fraction of potential signal positrons are deflected for lighter dark matter masses.
	}
	\label{fig:results:noB}
	\vspace*{-0.1in}
\end{figure}

\begin{figure}[h] 
	\hspace*{-.5cm} 
	\includegraphics[width=0.41\linewidth]{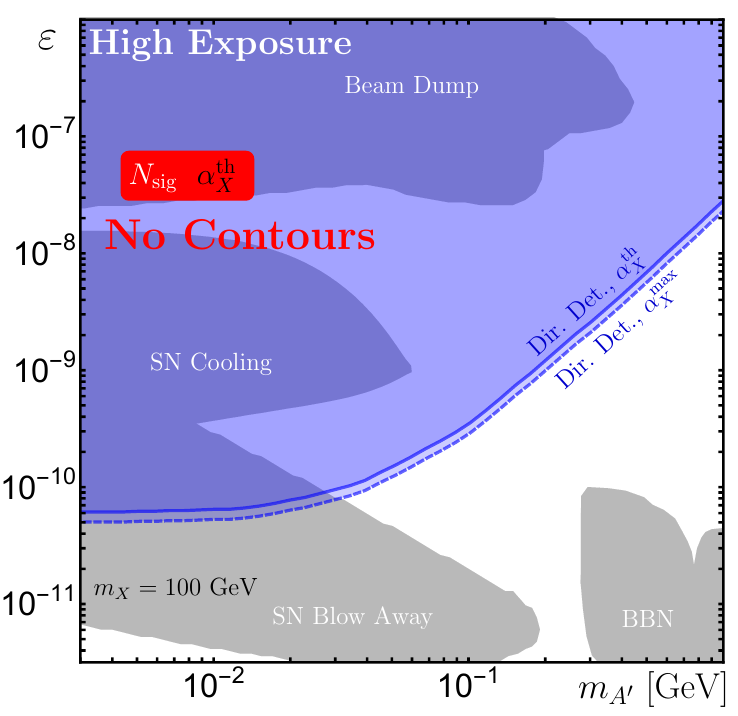} \qquad
	\includegraphics[width=0.41\linewidth]{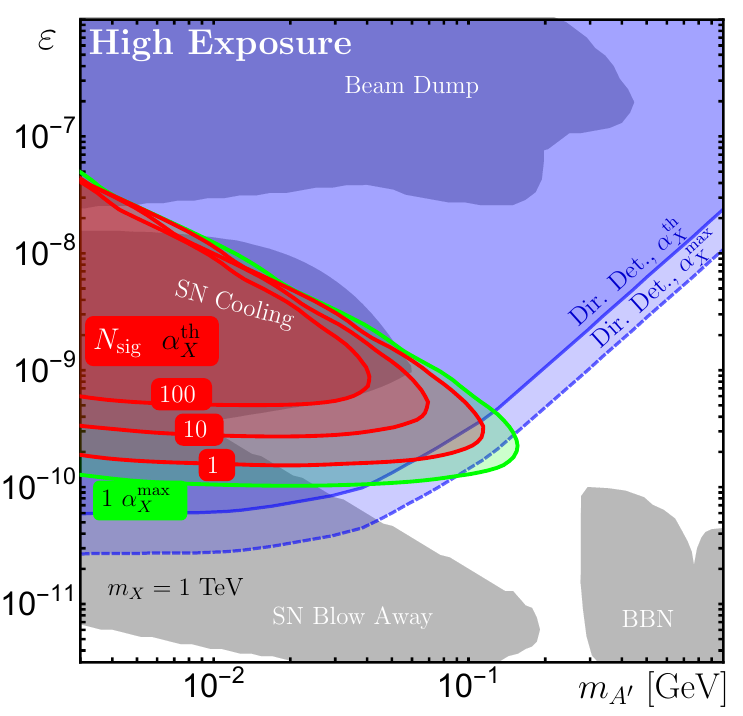} \\
	\includegraphics[width=0.41\linewidth]{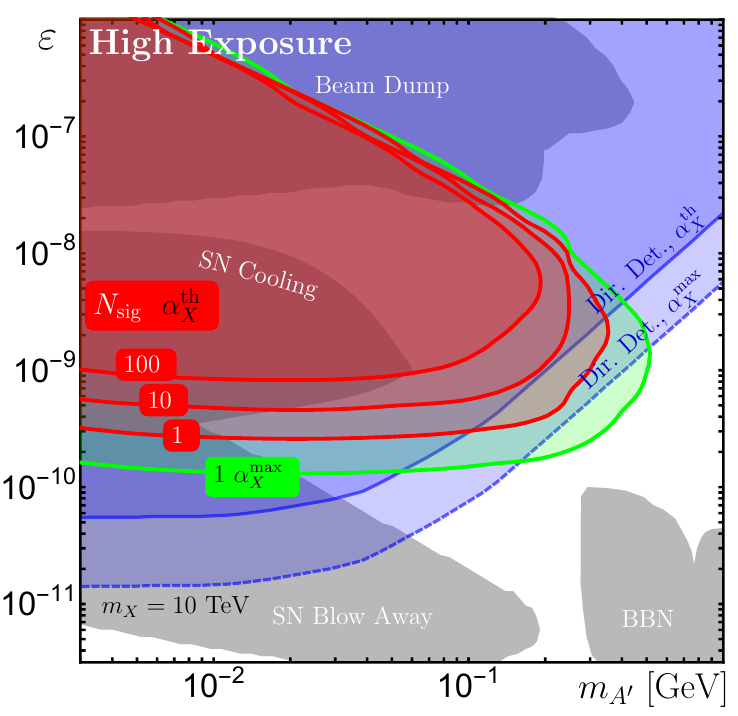} \qquad
	\includegraphics[width=0.41\linewidth]{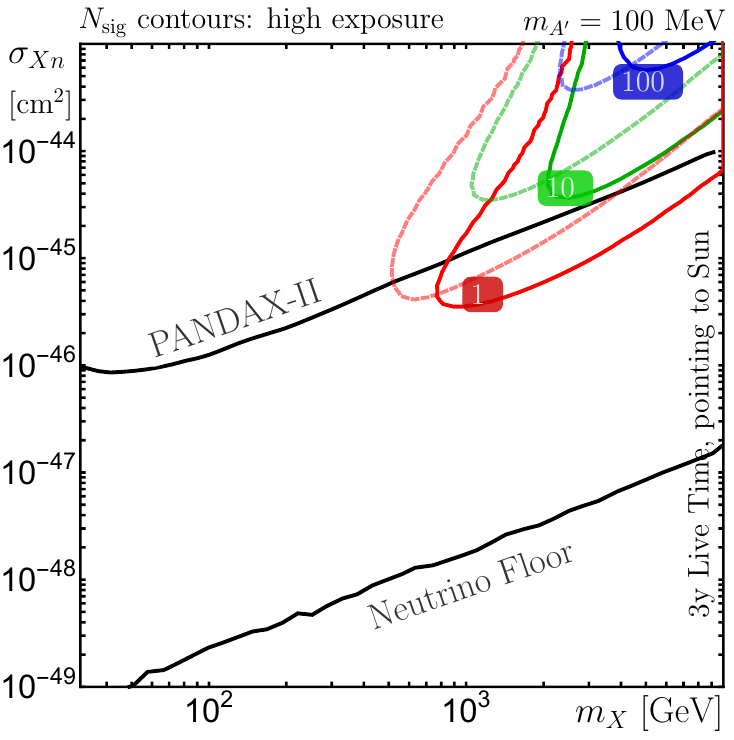} 
	\vspace*{-0.1in} 
	\caption{Same as Fig.~\ref{fig:results} but for a hypothetical high-exposure experiment with $\xi_\odot^\text{high} = 80 \xi_\odot$. The condition of a single background event $N_B=1$ in Eq.~(\ref{eq:theta:cut}) sets such strong cuts that there are no signal positrons for $m_X \lesssim 500~\text{GeV}$. 
	}
	\label{fig:results:hiX}
	\vspace*{-0.1in}
\end{figure}

The $N_S$ contours are shaped by the signal probability $\Pdet$ shown in Fig.~\ref{fig:acceptancediscussion} and described in \secref{signal}. This is in contrast to the case of Earth capture where the low-$\varepsilon$ portion of the contours were shaped by the equilibrium condition and followed the Sommerfeld resonances analogous to Fig.~\ref{fig:tausun}. Although the two scenarios are qualitatively similar, their signal reach is limited by different physics. For a fixed $m_X$, the search for dark photons from the Sun probes a region in the $(m_{A'},\varepsilon)$ plane probes a region below that of the Earth capture scenario presented in Ref.~\cite{Feng:2015hja}. This is as expected: solar dark photons must propagate further to escape the Sun than those from the Earth, and they thus provide sensitivity to a region of longer decay lengths $L$ and smaller $\varepsilon$.

The $N_S$ contours are not significance contours. A more detailed analysis is required to obtain significant contours, but we note that, in particular, in looking for an excess of signal positrons, we have treated all positrons with energies above $E_{\text{cut}}$ with equal weight.  This is a great oversimplification.  For example, for models with $m_X = 10~\tev$, the signal is optimized for $E_{\text{cut}} \sim 1~\tev$,  as seen in \figref{bestcuts10TeV}, and so any positrons from the Sun's direction with energy between around 1 and 10 TeV contributes to $N_S$.  But at the upper end of this range, the background is completely negligible, {\em even integrated over the whole sky}.   If AMS detected just one multi-TeV positron, and it came from the direction of the Sun, this would be quite significant.   In this case, the $N_S = 1$ contours may be thought of as characterizing the reach of AMS, whereas in other cases, requiring $N_S = 5$ over a background of $N_B = 1$ might be more reasonable.

With this caveat in mind, we now compare the signal reach to the sensitivities of other probes.  In \figref{results}, the 
dark photon bounds from colliders, beam dumps, and cosmology outlined in \secref{bounds} are shown in gray.
For dark matter masses $M_X \agt \tev$, the search reach extends well beyond these bounds---the latter in part due to the recent reanalysis in Ref.~\cite{Mahoney:2017jqk} which had found that prior estimates have overestimated the reach of these searches by about an order of magnitude. Even given collider experiment and cosmology bounds, AMS could detect tens or even hundreds of high energy positrons from the Sun.

Direct detection experiments are, however, more sensitive.  
Current bounds from PANDAX-II are also shown in \figref{results} in blue.  For the framework analyzed here, the AMS reach contours probe the same region of parameter space as existing direct detection searches. 
This is due, in part, to the solar magnetic field deflecting the positrons and smearing out what is otherwise a very clean directional signal for AMS. The severity of this effect can be seen by comparing to Fig.~\ref{fig:results:noB}, which shows the signal contours in the case where the solar magnetic field is ignored.

One may extend the signal reach by increasing the solar exposure. As a benchmark for this, Fig.~\ref{fig:results:hiX} shows the reach of a hypothetical `high solar exposure' experiment which the same properties as AMS but that points to the sun during its entire livetime. This corresponds to an exposure that is 80 times larger for $T=3$ years livetime~\cite{Machate}.

Direct detection experiments and the indirect detection signal analyzed here are, however, quite complementary.  As an example, in this paper we have focused on the case where dark matter scatters elastically.  However, the model already has all of the ingredients to introduce a pseudo-Dirac splitting between the dark matter states, if one assumes that the order parameter that controls the dark photon mass also gives a small Majorana mass to the $X$ and $\bar X$. This was most recently explored in Ref.~\cite{Izaguirre:2015zva} for collider searches of dark matter--dark photon systems. As is well known, only a modest splitting is required to suppress the direct detection signal~\cite{TuckerSmith:2001hy}.
In such a case, the solar capture process is largely unchanged. The splitting sets a lower bound on the relative velocity of dark matter--ordinary matter scattering, which sets an upper bound on the Sommerfeld enhancement.  However, since the Sun is a large enough target that it is in equilibrium through most of the relevant parameter space, this reduced Sommerfeld enhancement does not have a large effect on the dark matter annihilation rate. Thus it is simple to consider a regime in theory-space where the high-$m_X$ bounds in Fig.~\ref{fig:results} probe new territory. 
We emphasize that this regime does not require any new ingredients beyond the assumptions implicit in the benchmark model of this paper. We leave a detailed study of this scenario to future work. 

In the bottom--right panels of Figs.~\ref{fig:results}--\ref{fig:results:hiX}, we show these results in the usual direct detection plane $(m_X, \sigma_{Xn})$ where $\sigma_{Xn}$ is the $X$-nucleon cross section. We fix $m_{A'} = 100~\text{MeV}$.  The reach of the solar dark photon signal appears to be greater for $\alpha_X^\text{th}$ than for $\alpha_X^\text{max}$. This is because  $\sigma_{Xn} \sim \alpha_X \varepsilon^2$ so that $\sigma_{Xn}$ corresponds to a smaller value of $\varepsilon$ when assuming the maximal $\alpha_X^\text{max}$ dark sector coupling versus the thermal value $\alpha_X^\text{th}$.

\section{Conclusions} 

We have presented a novel method to discover dark sectors whose gauge bosons kinetically mix with the SM. Dark matter is captured by the Sun and can yield a smoking gun signature when it annihilates to dark photons that exit the Sun.  These dark photons then decay into $e^+ e^-$ pairs that may be searched for using directional discrimination from a space-based telescope such as AMS with its fantastic angular resolution. This search is insensitive to difficult-to-quantify astrophysical backgrounds and provides an opportunity for unambiguous dark matter discovery by AMS. 

We have presented a complete treatment in the dark photon scenario that includes several effects that had heretofore been neglected.
Our analysis incorporates the effect of non-perturbative Sommerfeld enhancements in the dark matter annihilation rate at the center of the Sun, which enlarges the region of parameter space in which dark matter capture and annihilation are in equilibrium. This is a necessary condition for a maximal annihilation rate. We have also addressed the non-perturbative enhancements in dark matter self-scattering at low velocities. These affect the rate of dark matter self-capture. In most of the phenomenologically relevant parameter space, self-capture remains a subdominant effect. We pointed out regimes that may be of interest for self-interacting dark matter models, where there may be significant deviations from our analysis.

We modeled the effects of the solar magnetic field on the experimental reach of the AMS detector. These magnetic fields smear out the signal, weakening the directionality of the signal, which would otherwise be effectively point-like.  Assuming high-energy positrons can be accurately ray-traced back to regions where the Earth's magnetic field is negligible, we defined a set of cuts that optimize the signal probability $\Pdet$ subject to a fixed number of allowed background events, and we estimated the reach for AMS with three years of data. The reach extends beyond regions probed by beam dump and supernova bounds, and is similar to the regions probed by direct detection.  These latter bounds, however, are much less stringent if the dark matter section includes even very small pseudo-Dirac mass splittings. Such splittings are generic in our framework and require no additional ingredients. We leave a detailed exploration of this scenario to future work \cite{Smolinsky:2017fvb}.  For comparison, we have also shown results for the case where the signal is not degraded by bending in a solar magnetic field, and for a hypothetical AMS-like experiment that points at the Sun and so has 80 times its exposure.  In both of these cases, again requiring negligible background, the number of signal events is improved by an order of magnitude.

In Ref.~\cite{Feng:2015hja} we showed that the IceCube experiment can be used to search for captured dark matter in the Earth annihilating into dark photons.  For dark sunshine leading to positrons and electrons, however, the IceCube signal is suppressed, since these positrons and electrons will be captured in the Earth before entering IceCube. However, if the dark photons decay into muons, these muons may penetrate through kilometers of earth to reach IceCube. Because the amount of earth between the Sun and IceCube is time dependent, this signal would have an annual modulation. Separately, we have shown in the appendix that gauge invariance requires dark photons to have a small coupling to the weak neutral current. For small masses this is suppressed relative to the coupling to the electric current, but such a neutrino signal would not be affected by the solar magnetic fields which afflict the positron signal. It may then be interesting to recast IceCube searches for solar neutrinos in terms of an excess coming from intermediate dark photons that decay to neutrinos.

\section*{Acknowledgments} 

We are grateful to Fabian Machate and
Stefan Schael for providing us with their results on the exposure of AMS to the Sun. We also
thank 
Pietro Baratella, 
Marco Cirelli, 
Eugenio Del Nobile,
Eder Izaguirre,
Gordan Krnjaic, 
Aldo Serenelli,
Brian Shuve,
Alex Wijangco,
and Hai-Bo Yu 
for helpful
discussions.
We thank Adam Green for pointing out a typo in our decay length code, which moved the region of experimental sensitivity to values of $\varepsilon$ that are lower by an order of magnitude.
J.L.F.~thanks Samuel Ting for the invitation to speak at and participate in ``AMS Days at CERN,'' which stimulated part of this work.
P.T.~thanks the Munich Institute for Astro- and Particle Physics
(MIAPP, DFG cluster of excellence "Origin and Structure of the
Universe") workshop ``Anticipating Discoveries: LHC14 and Beyond'' for
its hospitality and support during the completion of this work.
The work of J.L.F.\ and P.T.\ was performed in part at the Aspen Center for Physics, which is supported by National Science Foundation grant PHY--1066293.
This work is supported by NSF Grant No.~PHY--1316792.  
J.L.F. was supported in part by a Guggenheim Foundation grant and in
part by Simons Investigator Award \#376204.
P.T.~is supported in part by a UCI Chancellor's \textsc{advance} fellowship.
Numerical calculations were
performed using \emph{Mathematica 10.2}~\cite{Mathematica10}.

\appendix* 

\section{Diagonalization of the Dark Photon Hamiltonian}
\label{app:diagonalization}

Here we present a systematic derivation of the transformation from the dark photon gauge eigenstates to the mass eigenstates. The results in this appendix are known in the literature, see e.g.~\cite{Feldman:2007wj, Izaguirre:2015eya}; we present the derivation for clarification and to establish conventions. For simplicity, in this appendix we write the field strengths as $A'_{\mu\nu} = \partial_{[\mu}A'_{\nu]}$.

\subsection{Kinetic Mixing Between Massless and Massive Abelian Gauge Bosons}

We first examine the case of a massive U(1) gauge boson, $D$, mixing with a massless U(1) gauge boson, $B$. This is the diagonalization relevant for a dark photon ($D=A'$) which kinetically mixes with hypercharge in the limit $m_{D} \ll v$ so that the mixing is effectively only with the photon ($B=A$). The gauge-basis Lagrangian is
\begin{align}
\mathcal L &= 
-\frac 14 D_{\mu\nu}D^{\mu\nu}
-\frac 14 B_{\mu\nu}B^{\mu\nu}
+ \frac \varepsilon 2 B_{\mu\nu}D^{\mu\nu}
+ \frac 12 m_{D}^2 D_\mu D^\mu \ .
\label{eq:app:kin:mix}
\end{align}
We first remove the kinetic mixing term with a $\pi/4$ rotation,
\begin{align}
D &= 
\frac{D_1 - B_1}{\sqrt{2}}
&
B &= \frac{D_1 + B_1}{\sqrt{2}} \ ,
\label{eq:app:kin:mix:rot1}
\end{align}
where $B_1$ and $D_1$ are the rotated fields. The kinetic terms are now diagonal, but are not canonically normalized,
\begin{align}
\mathcal L &= 
-\frac 14 (1-\varepsilon) D_{1\mu\nu}D_1^{\mu\nu}
-\frac 14 (1+\varepsilon) B_{1\mu\nu}B_1^{\mu\nu}
+ \frac 14 m_{D}^2 (D_1 - B_1)_\mu (D_1 - B_1)^\mu \ .
\end{align}
To canonically normalize the kinetic terms, we perform a rescaling,
\begin{align}
D_1 &= \frac{D_2}{\sqrt{1 - \varepsilon}}
&
B_1 &= \frac{B_2}{\sqrt{1 + \varepsilon}}
\ .
\label{eq:app:easy:mixing:rescale}
\end{align}
With this, the kinetic terms are now universal and do not transform under subsequent rotations so that we are free to diagonalize the mass term. Were it not for the rescaling in Eq.~(\ref{eq:app:easy:mixing:rescale}), this would simply be a $-\pi/4$ rotation. Plugging in a general rotation,
\begin{align}
D_2 &= c_3 D_3 - s_3 B_3
&
B_2 &= s_3 D_3 + c_3 B_3
\ ,
\label{eq:app:kin:mix:rot3}
\end{align}
one finds that the mass matrix is diagonalized when\footnote{A shortcut to obtain this result is to observe that invariance of the unbroken U(1) gauge symmetry implies that the mass term for $B_3$ should also vanish. This coefficient of the mass matrix is simpler to solve than the off-diagonal element and gives $s_3 \propto \pm \sqrt{1-\varepsilon}$ and $c_3 \propto \mp \sqrt{1+\varepsilon}$.
}
\begin{align}
s_3 &= -\sqrt{\frac{1-\varepsilon}{2}}
&
c_3 &= \sqrt{\frac{1+\varepsilon}{2}}
\label{eq:app:kin:mix:rot3:params}
\ .
\end{align}
The choice of sign amounts to the sign of the $B$ coupling. Plugging this in gives the transformation from the gauge to energy eigenbasis:
\begin{align}
D &= \frac{1}{\sqrt{1-\varepsilon^2}}  D_3
&
B &= B_3 + 
\frac{\varepsilon}{\sqrt{1-\varepsilon^2}} D_3
\ .
\label{eq:app:diag:D:B}
\end{align}
From this we see that the dark photon picks up an $\mathcal O(\varepsilon)$ coupling to the $B$-current, $j_B\cdot B \supset \varepsilon_\text{eff} j_B\cdot D_3$, where $\varepsilon_\text{eff} = \varepsilon/\sqrt{1-\varepsilon^2}$, while the $B$ does not pick up any coupling to the dark current as expected by gauge invariance. The dark photon mass is rescaled to $m_D/\sqrt{1-\varepsilon^2}$.

For the case where $m_D = 0$, the gauge Lagrangian is diagonalized and normalized after Eq.~(\ref{eq:app:easy:mixing:rescale}) and the rotation in Eq.~(\ref{eq:app:kin:mix:rot3}) with parameters in Eq.~(\ref{eq:app:kin:mix:rot3:params}) is not strictly necessary. In fact, in this case one may chose to rotate the $D_2$ and $B_2$ into each other with any arbitrary SO(2) rotation. The choice in Eq.~(\ref{eq:app:kin:mix:rot3:params}) is convenient because it is close to the gauge basis. Phenomenologically, however, it is common to pick a rotation such that the ordinary photon couples to the dark current proportional to $\varepsilon$ so that the dark matter appears to be millicharged under electromagnetism. Ref.~\cite{Izaguirre:2015eya} calls this the Holdom phase. This interpretation is equivalent since in the case where $m_D$ is negligibly small, the photon and dark photon propagators are identical. Whether a process is identified as coming from a dark photon with $\varepsilon$ coupling to $j_\text{EM}$ or an ordinary photon with $\varepsilon$ coupling to $j_X$ is equivalent; and in general both diagrams must be included.

\subsection{Dark Photon--Hypercharge Mixing}

The dark photon--photon mixing is only an effective description since at high energies one must satisfy electroweak gauge invariance. This imposes that the UV mixing is actually between the dark sector U(1) and hypercharge, which is itself broken by the Higgs vev, $v$. Thus one must generically consider the mixing between the $D$, the photon $A$, and the $Z$ boson. The amount of $D$--$A$ mixing versus $D$--$Z$ mixing determines the extent to which the $D$ picks up the electroweak chiral couplings versus the vector-like electromagnetic couplings.

The hypercharge boson is related to the SM mass eigenstates by $B = -s_W Z + c_W A$. The mixing in Eq.~(\ref{eq:app:kin:mix}) is thus
\begin{align}
\frac{\varepsilon}{2} B_{\mu\nu}D^{\mu\nu}
&= 
-\frac{\varepsilon s_W}{2} Z_{\mu\nu} D^{\mu\nu}
+\frac{\varepsilon c_W}{2} A_{\mu\nu} D^{\mu\nu}
\ ,
\end{align}
with $A$ massless and $Z$ picking up an electroweak symmetry breaking mass of $M_Z$. The $D$--$A$ system is now identical to the $D$--$B$ system above, so we may diagonalize using the transformation in Eq.~(\ref{eq:app:diag:D:B}),
\begin{align}
D &= \frac{1}{\sqrt{1-\varepsilon^2 c_W^2}} D_1
&
A &= A_1 + \frac{\varepsilon c_W}{\sqrt{1-\varepsilon^2 c_W^2}} D_1
&
Z &= Z_1.
\label{eq:app:rot1}
\end{align}
This diagonalizes and canonically normalizes the $D$--$A$ system, but also changes the kinetic mixing between the $D$ and $Z$:
\begin{align}
-\frac{\varepsilon s_W}{2} Z_{\mu\nu} D^{\mu\nu}
&=
- \frac{1}{2} \frac{\varepsilon s_W}{\sqrt{1-\varepsilon^2 c_W^2}} Z_{1\mu\nu} D_1^{\mu\nu}
=
-\frac{\varepsilon_s}{2} Z_{1\mu\nu} D_1^{\mu\nu} 
\ .
\end{align}
In the above equation we have defined for convenience a new $D$--$Z$ mixing parameter $\varepsilon_s$
\begin{equation}
\varepsilon_s \equiv \frac{\varepsilon s_W}{\sqrt{1-\varepsilon^2 c_W^2}} \ .
\end{equation}
This kinetic mixing is removed with a $\pi/4$ rotation and canonical normalization is restored with a subsequent rescaling analogous to Eqs.~(\ref{eq:app:kin:mix:rot1},\ref{eq:app:easy:mixing:rescale}):
\begin{align}
D_1 &= \frac{D_2 - Z_2}{\sqrt{2}}
&
Z_1 &= \frac{D_2 + Z_2}{\sqrt{2}} 
\label{eq:app:rot2}
\\
D_2 &= \frac{D_3}{\sqrt{1+\varepsilon_s}}
&
Z_2 &= \frac{Z_3}{\sqrt{1-\varepsilon_s}}
\label{eq:app:rot3}
\ .
\end{align}
Unlike the previous case of a mixing between a massive and massless state, the $D$--$Z$ system is a mixing between two massive states. The mass matrix is diagonal in the ($D$, $A$, $Z$) basis where $D$ is a gauge eigenstate and $A$ and $Z$ are mass eigenstates with respect to the electroweak symmetry-breaking mass terms. Note that the $A$ has now decoupled completely and it is sufficient to consider the $D$--$Z$ system independently. For convenience, we perform a $-\pi/4$ rotation, which captures most of the rotation in Eq.~(\ref{eq:app:kin:mix:rot3}):
\begin{align}
D_3 &= \frac{D_4 + Z_4}{\sqrt{2}}
&
Z_3 &= \frac{Z_4 - D_4}{\sqrt{2}} 
\label{eq:app:rot4}
\ .
\end{align}
The original $D$ and $Z$ fields may now be written as 
\begin{align}
D &= a
\left(
\Delta D_4
+
\delta Z_4
\right)
&
Z &= 
\Delta Z_4 + \delta D_4
\ ,
\end{align}
where we define convenient shorthand,
\begin{alignat}{6}
a &= \frac{1}{\sqrt{1-\varepsilon^2 c_W^2}}
&&
=	1 + \frac 12 \varepsilon^2 c_W^2 + \mathcal O(\varepsilon^3)
\\
\Delta &= \frac{1}{2}\left(
\frac{1}{\sqrt{1+\varepsilon_s}} + 
\frac{1}{\sqrt{1-\varepsilon_s}}
\right)
&&
=
1 + \frac 38 \varepsilon_s^2 + \mathcal O(\varepsilon^3)
\\
\delta &= \frac{1}{2}\left(
\frac{1}{\sqrt{1+\varepsilon_s}} - 
\frac{1}{\sqrt{1-\varepsilon_s}}
\right)
&&
=
- \frac 12 \varepsilon_s+ \mathcal O(\varepsilon^4)
\ .
\end{alignat}
The mass term is then
\begin{align}
\frac 12
\begin{pmatrix}
D & Z
\end{pmatrix}
\begin{pmatrix}
m_D^2 & 
\\
& M_{Z}^2
\end{pmatrix}
\begin{pmatrix}
D \\ Z
\end{pmatrix}
&=
\frac 12
\begin{pmatrix}
D_4 & Z_4
\end{pmatrix}
\begin{pmatrix}
M_{11}^2 & M_{12}^2
\\
M_{12}^2 & M_{22}^2
\end{pmatrix}
\begin{pmatrix}
D_4 \\ Z_4
\end{pmatrix}
\ ,
\label{eq:app:mass:4}
\end{align}
where the elements on the right-hand side are, writing $\bar m_D^2 \equiv a^2 m_D^2$,
\begin{alignat}{3}
M_{11}^2 &  \;=\; 
\Delta^2  \bar m_D^2 
+ \delta^2 M_Z^2 
&&\;\sim\; 
m_D^2 +\mathcal O(\varepsilon^2)
\\
M_{22}^2 &\;=\; 
\Delta^2  M_Z^2 
+ \delta^2 \bar m_D^2 
&&\;\sim\; 
M_Z^2 +\mathcal O(\varepsilon^2)
\\
2 M_{12}^2 &\;=\;
2\delta \Delta
( \bar m_D^2 + M_Z^2 )
&&\;\sim\; 
-\varepsilon_s (m_D^2 + M_Z^2) + \mathcal O(\varepsilon^3) 
\ ,
\end{alignat}
One may now perform a final rotation to go the the mass eigenstates of the system. Observe that the off-diagonal element of the mass matrix is proportional to $\varepsilon$, so that the rotation is small in the small $\varepsilon$ limit. Writing $c=\cos\theta$ and $s=\sin\theta$, the rotation is given by 
\begin{align}
D_4 & =  c D_5 + s Z_5
&
Z_4 & = c Z_5 - s D_5
&
\tan 2 \theta &= \frac{2M_{12}^2}{M_{22}^2 - M_{11}^2}
\label{eq:app:rot5}
\ ,
\end{align}
where $c$ and $s$ are written to $\mathcal O(\varepsilon^2)$ as
\begin{align}
c &= 1 + \mathcal O(\varepsilon^2)
&
s &= - \frac{\varepsilon_s}{2} \frac{m_D^2 + M_Z^2}{M_Z^2-m_D^2} + \mathcal O(\varepsilon^3)
\ .
\end{align}
Plugging in the sequence of rotations in Eqs.~(\ref{eq:app:rot1}, \ref{eq:app:rot2}, \ref{eq:app:rot3}, \ref{eq:app:rot4}, \ref{eq:app:rot5}), the electroweak basis and mass basis are related by
\begin{align}
D &= \frac{\Delta c - \delta s}{\sqrt{1-\varepsilon^2c_W^2}} D_5
+ 
\frac{\Delta s + \delta c}{\sqrt{1-\varepsilon^2c_W^2}} Z_5
&
Z &= 
(\Delta c + \delta s) Z_5
+ (\delta c - \Delta s) D_5
\\
&= D_5 - \frac{\varepsilon_s M_Z^2}{M_Z^2 - \bar m_D^2}Z_5 + \mathcal O(\varepsilon^2) 
&
& = Z_5 + \frac{\varepsilon_s \bar m_D^2}{M_Z^2 - \bar m_D^2}D_5 + \mathcal O (\varepsilon^2)
\\
& = D_5 - \varepsilon_s Z_5 + \mathcal O(\varepsilon^2, \frac{m_D^2}{M_Z^2})
&
&= 
Z_5 + \varepsilon_s\frac{m_D^2}{M_Z^2} D_5 + \mathcal O(\varepsilon^2, \frac{m_D^2}{M_Z^2})
\ .
\label{eq:DD5:ZZ5} 
\end{align}
From this we observe that $D_5$ couples to the weak neutral current suppressed by $\varepsilon_s m_D^2/M_Z^2$, so that in the $m_D \ll m_Z$ limit the Standard Model couplings are effectively vector-like coming from the mixing with the photon. In this limit one may disregard the $D$--$Z$ mixing relative to the $D$--$A$ mixing.

\bibliography{bibdarkphotonAMS}

\end{document}